\documentclass[%
 reprint,
superscriptaddress,
 amsmath,amssymb,
 aps,
]{revtex4-2}

\usepackage[]{algorithm2e}
\usepackage{tikz}
\usetikzlibrary{shapes,arrows,positioning}
\tikzset{decision/.style={diamond, draw, fill=yellow!20, 
    text width=10em, aspect=2, text badly centered, node distance=3cm, inner sep=0pt, minimum height=4em},
block/.style={rectangle, draw, fill=gray!20, 
    text width=15em, text badly centered, rounded corners, minimum height=4em},
small_block/.style={circle, draw, fill=white!20, 
    text width=0em, text centered, rounded corners, minimum height=0em},    
line/.style={draw, -latex'},
cloud/.style={draw, ellipse,fill=red!20, node distance=3cm,text width=15em,
    minimum height=6em, text centered,},}

\usepackage[caption=false]{subfig}

\usepackage{graphicx}
\usepackage{dcolumn}
\usepackage{bm}
\usepackage{xcolor}
\usepackage{soul}
\usepackage{ulem}
\usepackage[hidelinks,colorlinks=true,linkcolor=blue,citecolor=blue]{hyperref}

\DeclareMathOperator{\Tr}{Tr}

\makeatletter
\let\save@mathaccent\mathaccent
\newcommand*\if@single[3]{%
  \setbox0\hbox{${\mathaccent"0362{#1}}^H$}%
  \setbox2\hbox{${\mathaccent"0362{\kern0pt#1}}^H$}%
  \ifdim\ht0=\ht2 #3\else #2\fi
  }
\newcommand*\rel@kern[1]{\kern#1\dimexpr\macc@kerna}
\newcommand*\wideaccent[2]{\@ifnextchar^{{\wide@accent{#1}{#2}{0}}}{\wide@accent{#1}{#2}{1}}}
\newcommand*\wide@accent[3]{\if@single{#2}{\wide@accent@{#1}{#2}{#3}{1}}{\wide@accent@{#1}{#2}{#3}{2}}}
\newcommand*\wide@accent@[4]{%
  \begingroup
  \def\mathaccent##1##2{%
    \let\mathaccent\save@mathaccent
    \if#42 \let\macc@nucleus\first@char \fi
    \setbox\z@\hbox{$\macc@style{\macc@nucleus}_{}$}%
    \setbox\tw@\hbox{$\macc@style{\macc@nucleus}{}_{}$}%
    \dimen@\wd\tw@
    \advance\dimen@-\wd\z@
    \divide\dimen@ 3
    \@tempdima\wd\tw@
    \advance\@tempdima-\scriptspace
    \divide\@tempdima 10
    \advance\dimen@-\@tempdima
    \ifdim\dimen@>\z@ \dimen@0pt\fi
    \rel@kern{0.6}\kern-\dimen@
    \if#41
      #1{\rel@kern{-0.6}\kern\dimen@\macc@nucleus\rel@kern{0.4}\kern\dimen@}%
      \advance\dimen@0.4\dimexpr\macc@kerna
      \let\final@kern#3%
      \ifdim\dimen@<\z@ \let\final@kern1\fi
      \if\final@kern1 \kern-\dimen@\fi
    \else
      #1{\rel@kern{-0.6}\kern\dimen@#2}%
    \fi
  }%
  \macc@depth\@ne
  \let\math@bgroup\@empty \let\math@egroup\macc@set@skewchar
  \mathsurround\z@ \frozen@everymath{\mathgroup\macc@group\relax}%
  \macc@set@skewchar\relax
  \let\mathaccentV\macc@nested@a
  \if#41
    \macc@nested@a\relax111{#2}%
  \else
    \def\gobble@till@marker##1\endmarker{}%
    \futurelet\first@char\gobble@till@marker#2\endmarker
    \ifcat\noexpand\first@char A\else
      \def\first@char{}%
    \fi
    \macc@nested@a\relax111{\first@char}%
  \fi
  \endgroup
}
\makeatother

\newcommand\doubleoverline[1]{\overline{\overline{#1}}}

\newcommand\widebarbar{\wideaccent\doubleoverline}

\begin{document}

\preprint{APS/123-QED}

\title{Optimal power extraction from active particles with hidden states}

\author{Luca Cocconi}
\email{luca.cocconi@crick.ac.uk}
\affiliation{The Francis Crick Institute, London NW1 1AT, United Kingdom}
\affiliation{Department of Mathematics, Imperial College London, South Kensington, London SW7 2BZ, United Kingdom}

\author{Jacob Knight}%
\affiliation{Department of Mathematics, Imperial College London, South Kensington, London SW7 2BZ, United Kingdom}

\author{Connor Roberts}
\affiliation{Department of Mathematics, Imperial College London, South Kensington, London SW7 2BZ, United Kingdom}

\date{\today}

\begin{abstract}

We identify generic protocols achieving optimal power extraction from a single active particle subject to continuous feedback control under the assumption that its spatial trajectory, but not its instantaneous self-propulsion force, is accessible to direct observation. Our Bayesian approach draws on the Onsager-Machlup path integral formalism and is exemplified in the cases of free run-and-tumble and active Ornstein-Uhlenbeck dynamics in one dimension. Such optimal protocols extract positive work even in models characterised by time-symmetric positional trajectories and thus vanishing informational entropy production rates. We argue that the theoretical bounds derived in this work are those against which the performance of realistic active matter engines should be compared.

\end{abstract}

\maketitle

Macroscopic living creatures such as horses and oxen have been utilised by humans for millennia to do useful work. A question of current theoretical and practical interest is the extent to which energy can be efficiently harvested from \emph{microscopic} active systems \cite{pietzonka2019autonomous,martinez2017colloidal,martinez2016brownian,speck2022efficiency,pumm2022dna,malgaretti2022}, whose motion is subject to non-negligible noise. The efficiency of existing many-particle microscopic active matter engines, such as turbines driven by the persistent motion of \textit{E.\ coli} bacteria in solution \cite{dileonardo2010bacterial, thampi2016active,ponisch2022pili}, is heavily limited by the difficulty of rectifying the incoherent motion of collections of individual swimmers with weak alignment interactions in the bulk. Even under idealised conditions, where individual active particles can be manipulated independently, strict upper bounds on extractable power are not well understood, particularly when only a subset of the observables characterising active motion are accessible to direct observation \cite{zhen2022optimal,berger2009optimal,esposito2012stochastic}. Here, we present a generic framework for the identification of protocols achieving optimal power extraction from a single active particle under continuous feedback control with the assumption that the instantaneous net velocity, $\dot{x}(t)$, but \emph{not} the fluctuating contribution originating from the self-propulsion, $w(t)$, is observable. This is typically the case for realistic active matter engines \cite{pietzonka2019autonomous,dileonardo2010bacterial}. Our Bayesian approach, which draws on the Onsager-Machlup path integral formalism \cite{onsager1953fluctuations}, applies to a generic stochastic self-propulsion process and is illustrated in the cases of free run-and-tumble (RnT) \cite{garcia2021run} and active Ornstein-Uhlenbeck (AOU) \cite{martin2021statistical} dynamics in one dimension. 

Both models are characterised by time-symmetric positional trajectories (SM Sec.~\ref{sr:I}) and thus vanishing informational entropy production rates (iEPR) \cite{fodor2022irreversibility,markovich2021thermodynamics}, defined as 
the Kullback-Leibler divergence \cite{kullback1951information} per unit time of the ensemble of forward paths and their time-reversed counterparts \cite{gaspard2004time,roldan2015decision}.

In the Markovian case, where all degrees of freedom are observable, the iEPR is proportional to the thermodynamic dissipation and thus provides a (loose) upper bound to the extractable power. This relation fails to apply in the presence of hidden states \cite{cocconi2022scaling,yu2021inverse,esposito2012stochastic}. Indeed, we show that positive average power extraction remains possible even for vanishing iEPR upon Bayesian inference of the hidden state (cf.~\cite{PhysRevLett.129.220601}, where it is argued that vanishing local iEPR implies zero extractable work). Measurement-driven protocols of the type we discuss in the following incur a thermodynamic maintenance cost \cite{leff2002maxwell,ouldridge_power_2018}, but are not constrained by Landauer's principle in the same way as equilibrium information engines \cite{malgaretti2022,saha2022information}.

\paragraph*{Definition of the optimal protocol ---}  
Consider the overdamped Langevin equation for a generic active particle $\dot{x}(t) = w(t) + \gamma^{-1} F_{\rm ext}(t) + \sqrt{2 D_x} \xi(t)$, where $\xi(t)$ is a white noise of unit covariance with associated diffusivity $D_x$ and $\gamma$ denotes the viscosity. We henceforth work in units whereby $\gamma=1$. Here, $w(t)$ is a stochastic self-propulsion velocity, which for the time being we take to be measurable by an external observer tasked with controlling the applied force $F_{\rm ext}(t)$. {In practice, $F_{\mathrm{ext}}(t)$ could be implemented using an optical trap \cite{saha2022information} or, for a charged active colloid \cite{sandoval2016magnetic}, through an external electric field of time-varying magnitude and direction.} Positive average work is readily extracted by applying an $F_{\rm ext}(t)$ smaller than and opposite to the particle's self-propulsion \cite{pietzonka2019autonomous, dileonardo2010bacterial, roberts2023run}. Over a duration $T$ this generates a noise-averaged total work by the particle against the known external force
\begin{align}
\label{eq:work_definition}
    \mathbb{E}_\xi[ W_{\rm tot}[F_{\rm ext}] ] &= - \int_0^T dt \ F_{\rm ext}(t) \mathbb{E}_\xi[\dot{x}(t)] \nonumber\\
    &= - \int_0^T dt \ F_{\rm ext}(t)(w(t) + F_{\rm ext}(t))~,
\end{align}
{which constitutes the key observable of a hypothetical experiment.} Above and henceforth, $\mathbb{E}_{\phi}[\bullet]$ is used to denote an average with respect to the steady-state distribution of the random variable $\phi$. We will subsequently refer to $F_{\rm ext}(t)$ as ``the protocol''. The integrand of Eq.~(\ref{eq:work_definition}), corresponding to the instantaneous power output, can be maximised at each time $t$ by applying the protocol $F_{\rm ext}^*(t) = - w(t)/2$. The corresponding steady-state average power output is
\begin{equation} \label{eq:opt_w_colorful}
    \lim_{T \to \infty} \frac{\mathbb{E}_\xi[ W_{\rm tot}[F_{\rm ext}] ]}{T} = \frac{\bar{w}^2}{4} + \frac{\mathbb{E}_w[ (w(t)-\bar{w})^2 ]}{4}~,
\end{equation}
where $\bar{w} \equiv \mathbb{E}_w[ w(t)]$ and we have invoked ergodicity to convert time averages to ensemble averages. The average power is smaller than the thermodynamic dissipation at $F_{\rm ext} =0$, given by $D_x \dot{S}_i = \mathbb{E}_w[ w^2(t) ]$ \cite{seifert2012stochastic, cocconi2020entropy}, demonstrating that the entropy production rate $\dot{S}_i$ provides only a \textit{loose} upper bound to the extractable power at low Reynolds number, due to the unavoidability of viscous effects when $\dot{x}(t)\neq 0$. We will henceforth refer to protocols $F_{\rm ext}^*(t)$ achieving the maximum average power output allowed under a particular set of constraints as \emph{optimal}. 

\begin{figure}
    \centering
    \includegraphics[scale=0.215, trim = 0.8cm 0.4cm 0.6cm 0.4cm, clip]{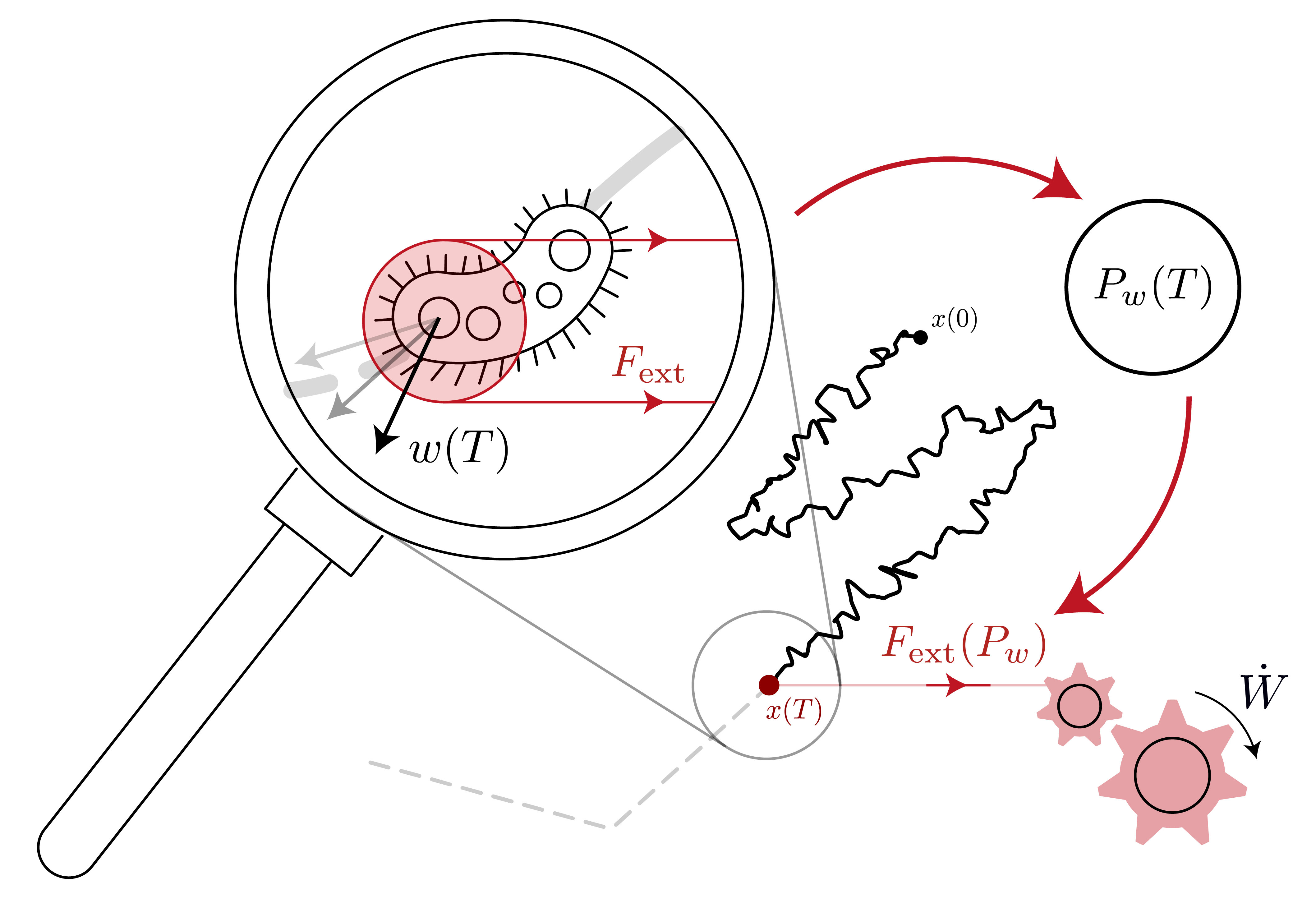}
    \caption{Optimal power extraction from an active particle (here visualised as a bacterium) with hidden self-propulsion velocity is achieved by subjecting the latter to continuous feedback control, whereby the magnitude and direction of the protocol $F_{\rm ext}(t)$ are modulated according to the inferred self-propulsion velocity. }
    \label{fig:schematic}
\end{figure}

Consider now the case where the underlying dynamics of the active particle (in the form of the full set of governing equations) are known but the instantaneous self-propulsion velocity $w(t)$ is not accessible to direct observation, i.e.\ it is a hidden variable. Na\"{i}vely, this suggests positive work extraction is unattainable since there is no immediate indication of which direction and magnitude should be chosen for $F_{\rm ext}(t)$. However, since $w(t)$ can still be partially inferred from the history of $x(t)$, positive work can be extracted during transient periods of persistent motion. To see this, let $\mathbb{P}(w(T)=v|\{x\}_0^T)$ denote the posterior probability density that the instantaneous self-propulsion velocity of the active particle at current time $T$ equals $v$ \emph{given} a particular spatial trajectory $\{x\}_0^T$ has been observed. 
The expected work extracted during a time window of duration $T$ can be expressed as the following functional of the generic protocol $F_{\rm ext}(t)$,
\begin{align}
    &\mathbb{E}_{\xi,w} [W_{\rm tot}[F_{\rm ext}] ] \nonumber \\
    &= - \int_0^T dt \int_{- \infty}^\infty dv \  \mathbb{P}(v|\{x\}_0^t)  F_{\rm ext}(t)(v + F_{\rm ext}(t))~.  \label{eq:mean_work}
\end{align}
The optimal protocol $F_{\rm ext}^*(t)$ is obtained 
from $\left. \delta \mathbb{E}_{\xi,w} [W_{\rm tot}[F_{\rm ext}] ]/\delta F_{\rm ext} \right|_{F_{\rm ext}^*} = 0$,
whence
\begin{equation}\label{eq:opt_force_general}
    F_{\rm ext}^*(T) = - \frac{1}{2} \int_{-\infty}^\infty dv \ \mathbb{P}(v|\{x\}_0^T) v = - \frac{\mathbb{E}_w[ w (T) | \{x\}_0^T ]}{2}~,
\end{equation}
where $\mathbb{E}_w[ w (T) | \{x\}_0^T ]$ denotes the posterior expectation of the self-propulsion velocity with respect to $\mathbb{P}(v|\{x\}_0^T)$. This is not to be confused with the expectation of $w(T)$ taken with respect to the corresponding prior probability $\mathbb{P}(v) = \int \mathcal{D}x~ \mathbb{P}(v|\{x\}_0^T) \mathbb{P}(\{x\}_0^T)$, which we denoted $\bar{w}$ and assume to be independent of $T$. Substituting the optimal force into the expression for the instantaneous power output, the integrand in Eq.~\eqref{eq:mean_work} gives
\begin{equation} \label{eq:opt_w_colorless}
    \mathbb{E}_{\xi,w}[ \dot{W}[F_{\rm ext}^*(t)] ] = \frac{\bar{w}^2}{4} + \frac{\mathbb{E}_w[ (w (T) - \bar{w}) | \{x\}_0^T ]^2}{4}~,
\end{equation}
cf.\ Eq.~\eqref{eq:opt_w_colorful}. In the following, we take $\bar{w}=0$ to focus on the non-trivial term appearing on the right-hand side of Eq.~\eqref{eq:opt_w_colorless}. Figure~\ref{fig:schematic} schematises the feedback control described above. 

\paragraph*{Warm-up: The run-and-tumble particle --- } 
We have reduced the problem of identifying the optimal protocol to the evaluation of the posterior expectation $\mathbb{E}_w[ w (T) | \{x\}_0^T ]$, Eq.~\eqref{eq:opt_force_general}. Now we proceed to show how this can be done for the case of RnT motion in one dimension, $\dot{x}(t) = \nu w(t) + F_{\rm ext}(t) + \sqrt{2D_x} \xi(t)$, whose binary internal self-propulsion mode $w(t)$ constitutes the simplest example of a state-space amenable to non-trivial coarse graining. 

In particular, let $w(t) \in \{-1,1\}$ be a dimensionless dichotomous noise with symmetric transition rate $\alpha$. We seek the posterior probability that the particle is a right self-propeller, $w(T)=+1$, given its positional trajectory up to the current time $T$, which we denote $P_+(T) = \mathbb{P}[w(T) = +1|\{x\}_0^T]$ for compactness. The complementary probability is denoted $P_-(T) = \mathbb{P}[w(T) = -1|\{x\}_0^T]$. Defining the \emph{confidence parameter} $Q[\{x\}_0^T] = \log(P_+(T)/P_-(T))$ and using $P_+(T) + P_-(T) = 1$, we can write
\begin{equation} \label{eq:def_Q}
    P_+(T) = \frac{e^Q}{1+e^Q} = \frac{1}{2}  + \frac{e^Q - 1}{2(1+e^Q)}~.
\end{equation}
Equation~(\ref{eq:def_Q}) reduces to the prior probability $\mathbb{P}(w=\pm 1) = 1/2$ when $Q=0$. To calculate $P_+(T)$ via $Q$ we thus need to find an expression for the ratio of the conditional path probabilities. To do so, we first invoke Bayes' theorem,
\begin{equation}
    \mathbb{P}[w(T) = \pm1|\{x\}_0^T] = \frac{\mathbb{P}[\{x\}_0^T | w(T) = \pm 1]}{2 \mathbb{P}[\{x\}_0^T]}~,
\end{equation}
where we have used $\mathbb{P}[w(T)=\pm 1]=1/2$. We can equivalently write
\begin{equation} \label{eq:s_rev_bay}
    Q[\{x\}_0^T] = \log \frac{\mathbb{P}[\{x\}_0^T | w(T) = + 1]}{\mathbb{P}[\{x\}_0^T | w(T) = - 1]}~,
\end{equation}
reminiscent of a stochastic entropy \cite{roldan2015decision}. We now introduce the notation for the average with respect to the distribution of $w(t)$ path probabilities conditioned on a particular final value $w(T)$,
\begin{equation} \label{eq:uparrow_notation}
\overline{\bullet}^{(v)} \equiv \int \mathcal{D}w \ \bullet \mathbb{P}[ \{w(t)\}_0^T| w(T)=v]~,
\end{equation}
which allows us to express the path probabilities in Eq.~\eqref{eq:s_rev_bay} as
\begin{subequations}\label{eq:cond_path_probs}
\begin{alignat}{2}
    \mathbb{P}[\{x\}_0^T | w(T) = + 1] &= \overline{\mathbb{P}[\{x\}_0^T | \{w\}_0^T] }^{(+1)}~, \label{eq:cond_path_pr} \\
    \mathbb{P}[\{x\}_0^T | w(T) = - 1] &= \overline{\mathbb{P}[\{x\}_0^T | \{w\}_0^T] }^{(-1)}~. 
\end{alignat}
\end{subequations}
Finally, we invoke the Onsager-Machlup path integral form \cite{onsager1953fluctuations} of the conditional path probability in the Stratonovich discretisation 
\begin{align} \label{eq:ons_mach_rnt}
    \mathbb{P}[\{x\}_0^T | \{w\}_0^T]
    \propto \exp\left( -\frac{1}{4D_x} \int_0^T dt \ \left(\dot{x}_c(t) - \nu w(t)\right)^{2} \right)~,
\end{align}
where $\dot{x}_c = \dot{x} - F_{\rm ext}$ denotes the velocity in the reference frame where the externally imposed drift is subtracted away. Substituting Eq.~\eqref{eq:ons_mach_rnt} into Eq.~\eqref{eq:cond_path_probs}, combining the resulting expressions with Eq.~\eqref{eq:s_rev_bay}, and cancelling common $w(t)$-independent factors appearing in the numerator and denominator, we eventually arrive at
\begin{align}
    Q[\{x\}_0^T] 
    = &\log \left(~ \overline{\exp \left( \frac{\nu}{2D_x} \int_0^T dt \ 
    \dot{x}_c(t)w(t) \right)}^{(+1)}~ \right) \nonumber \\
    &- \log \left(~ \overline{\exp \left( \frac{\nu}{2D_x} \int_0^T dt \ \dot{x}_c(t)w(t) \right)}^{(-1)} ~\right)  \label{eq:Q_exact}~,
\end{align}
where we have also used $w^2(t) = 1$ for all $t \in [0,T]$. To make further progress we exploit the identity between the logarithm of a moment-generating function and its cumulant-generating function \cite{lebellac1991,bothe2022particle}, as well as the parity of the cumulants (see SM Sec.~\ref{sr:II}). This leads to 
\begin{equation} \label{eq:Q_gen_rnt}
    Q[\{x\}_0^T] = \sum_{n \ \text{odd}}^\infty \frac{{\rm Pe}^n}{2^{n-1} n!} \overline{Y^n[\{x\}_0^T]}^{(+1),c} ,
\end{equation}
with P\'{e}clet number ${\rm Pe} = \nu^2/(D_x \alpha)$ and
\begin{equation}
    Y^n[\{x\}_0^T] = \int_0^T dt_1 ... dt_n  \prod_{i=1}^n \left( \frac{\dot{x}_c(t_i) \alpha}{\nu} \right) w(t_i)~,
\end{equation}
where the superscript $c$ in expectations, e.g.\ $\overline{\bullet}^{(v),c}$, denotes the corresponding cumulant. Substituting Eq.~\eqref{eq:Q_gen_rnt} into Eq.~\eqref{eq:def_Q}, combined with Eq.~\eqref{eq:opt_force_general}, returns the optimal protocol. 

Computing the right-hand side of Eq.~\eqref{eq:Q_gen_rnt} is unfeasible in general. However, $Q[\{x\}_0^T]$ can be computed analytically in the low-$\rm {Pe}$ asymptotic regime. To leading order in ${\rm Pe} \ll 1$, only the first cumulant $\overline{Y[\{x\}_0^T]}^{(+1),c}$ is required, which in turn draws on $\overline{w(t)}^{(+1),c} = \overline{w(t)}^{(+1)} = \exp(-2 \alpha(T-t))$, SM Sec.~\ref{sr:II}, whence we find
\begin{equation}\label{eq:Q_leading}
    Q[\{x\}_0^T] = {\rm Pe} \int_0^T dt \ \left( \frac{\alpha \dot{x}_c(t)}{ \nu} \right) e^{-2 \alpha(T-t)} + \mathcal{O}\left( {\rm Pe}^3 \right)~.
\end{equation}
In order to conveniently apply the optimal protocol under continuous feedback control, we can differentiate Eq.~\eqref{eq:Q_leading} with respect to $T$ and use the Leibniz integration rule (assuming $\dot{x}_c(t) = 0$ for $t<0$) to obtain a differential equation for the time evolution of $Q$, i.e.\ $\dot{Q}(T) = \nu \dot{x}_c(T)/D_x - 2 \alpha Q(T)$. Remarkably, upon substituting for $\dot{x}_c$ and rescaling time by the switching rate, $T' = \alpha T$, the Langevin equation for $Q(T')$ reads like that of a RnT particle in a harmonic potential with self-propulsion speed and diffusivity both equal to the P\'{e}clet number, i.e.\ 
\begin{equation} \label{eq:Qeq_resc}
    \frac{d Q(T')}{dT'} = {\rm Pe} \ w(T') - 2 Q(T') + \sqrt{2 {\rm Pe}} \ \xi(T')~.
\end{equation}

We now proceed to make the connection with the rate of work extraction. First of all, we have by combining Eq.~\eqref{eq:opt_force_general} and Eq.~\eqref{eq:def_Q} that the optimal protocol is given to leading order in $Q \sim {\rm Pe}$ by $F_{\rm ext}^*(T) = - \frac{\nu}{4} Q + \mathcal{O}(Q^2)$. When the optimal protocol is applied at all times, the resulting noise-averaged power output, Eq.~\eqref{eq:opt_w_colorless}, is given by $\mathbb{E}_{\xi}[ \dot{W}_{\rm RnT}[F_{\rm ext}^*(t)] ] = \nu^2 Q^2(T)/16 + \mathcal{O}\left({\rm Pe}^2\right)$. Taking a further expectation with respect to the dichotomous noise $w(t)$ and exploiting the mapping of the $Q$ dynamics onto those of a RnT particle in a harmonic potential, Eq.~\eqref{eq:Qeq_resc}, whence $\mathbb{E}_{\xi,w}[Q^2] = \left( 1 + {\rm Pe}/4 \right)${\rm Pe}/2 \cite{garcia2021run}, we eventually arrive at 
\begin{equation} \label{eq:RnT_lowPe_asym}
    \mathbb{E}_{\xi,w}[ \dot{W}_{\rm RnT}(F^*_{\mathrm{ext}})] = \frac{\nu^2}{4} \frac{\rm Pe}{8} + \mathcal{O}\left( {\rm Pe}^2 \right) ~,
\end{equation}
which constitutes a tight upper bound to the average extractable power from a RnT particle with hidden self-propulsion velocity in the low-${\rm Pe}$ regime. Higher moments of the fluctuating power output under $F_{\rm ext}^*$ can be computed similarly, see SM Sec.~\ref{sr:III}.

\begin{figure}
    \centering
    \includegraphics[width=0.97\columnwidth, trim = 0.3cm 0.5cm 0.4cm 0.2cm, clip]{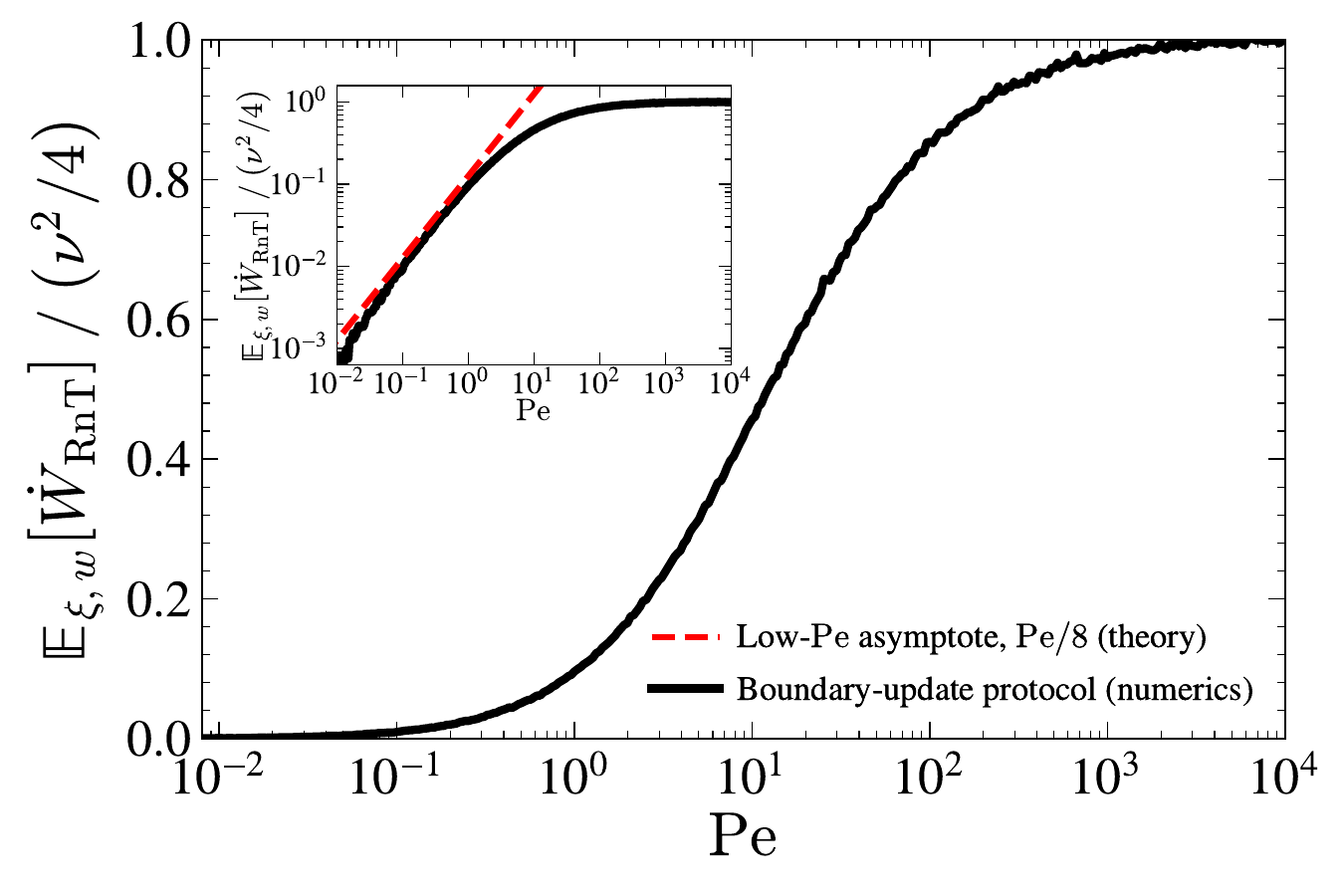}
    \caption{Average power extracted from a RnT particle with hidden self-propulsion velocity upon application of the boundary-update protocol, the numerical implementation of which is discussed in detail in SM Sec.~\ref{sr:IV}. The extractable power, which is positive for all $\rm Pe$, asymptotically approaches that of a situation where the internal state is known, Eq.~(\ref{eq:opt_w_colorful}), as $\rm Pe \rightarrow \infty$ and is in excellent agreement with the theoretical bound in the low-$\mathrm{Pe}$ limit, Eq.~(\ref{eq:RnT_lowPe_asym}).}
    \label{fig:BoundaryProtocol}
\end{figure}

\paragraph*{A boundary-update protocol --- } We further introduce an independent approach to computing the posterior probability $P_+(T)$ in real time. This novel ``boundary-update'' protocol, described in full detail in SM Sec.~\ref{sr:IV}, both saturates the bound \eqref{eq:RnT_lowPe_asym} and is conjectured to achieve optimality for \emph{all} $\rm Pe$. It draws on the \emph{conditional} splitting probabilities of the RnT process, which, to the best of our knowledge, we compute here for the first time. These are the probabilities that a particle initialised at $x_0 \in \left[-L/2, L/2 \right]$ in a given statistical superposition of internal states exits said interval through either the left or right boundary in either a left or right self-propulsion state. Knowledge of the splitting statistics is used in combination with Bayes' theorem to update the posterior distribution of the internal state $w(t)$ each time the particle is observed to undergo a net displacement larger than $L/2$ in the reference frame where the deterministic drift is subtracted away, $\dot{x}_c=\dot{x} - F_{\rm ext}$. In the limit $L \rightarrow 0$, the posterior updating frequency diverges and we conjecture that optimal inference is achieved. Figure~\ref{fig:BoundaryProtocol} shows application of the boundary-update approach indeed produces an average power output matching the bounds Eqs.~\eqref{eq:RnT_lowPe_asym} and \eqref{eq:opt_w_colorful} in the low- and high-${\rm Pe}$ limits, respectively.

\paragraph*{A generic active particle ---}
Having explored the particular case of RnT motion in some detail, we now expand our scope to a one-dimensional active particle with self-propulsion velocity $w(t)$ evolving according to a generic (discrete- or continuous-state) stochastic process \cite{caprini2022parental}.  
Following Eq.~\eqref{eq:opt_force_general}, the identification of the optimal protocol requires us to compute the posterior expectation of the self-propulsion velocity, which can be conveniently expressed as
\begin{align} \label{eq:generic_vexpect}
   &\mathbb{E}_w[ w(T) | \{x\}_0^T] 
    = \Tr_v \left[v  \mathbb{P}[w(T)=v | \{x\}_0^T] \right] \nonumber \\
    &=
    \frac{\Tr_v \left[ v \cdot \overline{{\rm exp}\left( - \frac{\rm Pe}{4} \int_0^T dt \frac{\mu}{\sigma_w^2} (\dot{x}_c(t)-w(t))^2 \right)}^{(v)} \mathbb{P}(v) \right]}{\Tr_v \left[ \overline{{\rm exp}\left( - \frac{\rm Pe}{4} \int_0^T dt \frac{\mu}{\sigma_w^2} (\dot{x}_c(t)-w(t))^2 \right)}^{(v)} \mathbb{P}(v) \right] }
\end{align}
with ${\rm Pe} = \sigma_w^2/(\mu D_x)$, $\sigma_w^2 = \mathbb{E}_w[ w^2]$, and $\mu$ a characteristic inverse timescale associated with the self-propulsion dynamics. 
Here, $\Tr_v$ denotes an integral (sum) over the continuous (discrete) state space.
We have also invoked Bayes' theorem to write
\begin{align}
    \mathbb{P}[w(T)=v | \{x\}_0^T]
    &= \frac{\mathbb{P}(v)  \overline{\mathbb{P}[ \{x\}_0^T | \{w(t)\}_0^T ]}^{(v)}}{\mathbb{P}[ \{x(t)\}_0^T]} ~,
\end{align}
where $\overline{\bullet}^{(v)}$ is defined as in  Eq.~\eqref{eq:uparrow_notation}, and we have used the normalisation condition $1 = \Tr_v  \mathbb{P}[w(T)=v | \{x\}_0^T]$ to divide by a factor of unity throughout, producing the same type of cancellations of $v$-independent terms observed in the RnT case. We can rewrite Eq.~\eqref{eq:generic_vexpect} in a compact form as 
\begin{equation} \label{eq:v_opt_gen_L}
    \mathbb{E}_w[ w(T) | \{x\}_0^T]
    = \frac{\Tr_v \left[ v \cdot e^{\mathcal{L}[\{x\}_0^T,v]} \mathbb{P}(v) \right]}{\Tr_v \left[ e^{\mathcal{L}[\{x\}_0^T,v]} \mathbb{P}(v) \right]}
\end{equation}
by introducing the cumulant-generating function
\begin{small}
\begin{align}
    \mathcal{L}[\{x\}_0^T,v] 
    &= \sum_{n=1}^\infty \frac{(- {\rm Pe})^n}{2^{2n} n!} \overline{ \left[ \int_0^T dt \ \frac{\mu}{\sigma_w^2} (\dot{x}_c(t)-w(t))^2 \right]^n \ }^{(v),c}~. \label{eq:L_expansion_gn}
\end{align}
\end{small}If no further assumptions can be made regarding the process $w(t)$, one can now truncate the sum and substitute the resulting expression into Eq.~\eqref{eq:v_opt_gen_L} to obtain, by invoking Eq.~\eqref{eq:opt_force_general} and recalling $\mathbb{E}_w[w]=0$, the optimal protocol in the asymptotic case ${\rm Pe} \ll 1$, 
\begin{small}
\begin{equation}\label{eq:mathcalL_exp_one}
    F^*_{\rm ext}(T)
    = \Tr_v \left[ v \frac{\rm Pe}{8 } \int_0^T dt \frac{\mu}{\sigma_w^2} \Big( \overline{w^2(t)}^{(v),c}- 2\dot{x}_c(t) \overline{w(t)}^{(v),c} \Big) \mathbb{P}(v) \right]~.
\end{equation}
\end{small}The form of Eq.~\eqref{eq:mathcalL_exp_one} matches the RnT result, Eq.~\eqref{eq:Q_leading}, except for the appearance of a term depending on the second-order cumulant $\overline{w^2(t)}^{(v),c}$, which was absent in the RnT case due to the norm of the self-propulsion velocity being constant. The correlation functions of the hidden state $w(t)$ in Eq.~\eqref{eq:mathcalL_exp_one} can be reconstructed from observable trajectories (see SM Sec.~\ref{sr:new_infer}), allowing us to relax \emph{a posteriori} the requirement that the equations governing the dynamics of $w(t)$ be known, and to extract work even in this case.

In SM Sec.~\ref{sr:V}, we apply the general result obtained above to the specific case of a one-dimensional AOU process, the simplest canonical active particle model with a continuous self-propulsion state \cite{martin2021statistical}. We find the average extractable power from an AOU particle with hidden self-propulsion velocity in the low-${\rm Pe}$ asymptote is bound above by 
\begin{equation} \label{eq:asym_lowPe-Aou}
    \mathbb{E}_{\xi,w}[\dot{W}_{\rm AOU}(F^*_{\rm ext})] = \frac{\sigma_w^2 }{4} \frac{\rm Pe}{16} + \mathcal{O}({\rm Pe}^2)
\end{equation}
and further compute the second moment of the power output distribution (SM Sec.~\ref{sr:III}). 

\paragraph*{ Langevin dynamics: high-Pe asymptotics ---}
When the dynamics of $w(t)$ are described by a Langevin process, Eq.~\eqref{eq:generic_vexpect} also allows us to explore the high-${\rm Pe}$ asymptote through a saddle-point expansion. For the particular case of the AOU process (as defined in SM Sec.~\ref{sr:V}), we can write, using the Onsager-Machlup form of $\mathbb{P}[\{w\}_0^T]$,
\begin{align}
    &\overline{{\rm exp}\left( - \frac{\rm Pe}{4} \int_0^T dt \frac{\mu}{\sigma_w^2} (\dot{x}_c(t)-w(t))^2 \right)}^{(v)} \nonumber \\
    &~~~~~~~~~~~~~~\propto  \int \mathcal{D}w \ e^{ -       \mathcal{N}[w(t);\{x\}_0^T] } \ \delta(w(T)-v)~, \label{eq:two_costs}
\end{align}
with the action-like functional
\begin{align} 
    \mathcal{N}[w(t);\{x\}_0^T]  = \mu \int_0^T dt \Bigg[   {\rm Pe} \left( \frac{\dot{x}_c(t)-w(t)}{2 \sigma_w} \right)^2 & \nonumber \\
    +  \left( \frac{\dot{w}(t)/\mu + w(t)}{2 \sigma_w}  \right)^2& \Bigg]~,\label{eq:action_func_m}
\end{align} 
which combines a ``potential'' term (prefactor $\rm Pe$), penalising departures from $w(t)=\dot{x}_c$, and a ``kinetic'' term (unit prefactor) penalising changes in $w(t)$ that are exceedingly fast or slow compared to the characteristic inverse timescale $\mu$ of the self-propulsion dynamics. Even at high ${\rm Pe}$, the second term cannot be ignored since the boundary condition $w(T)=v$ in general prevents $w(t)=\dot{x}_c(t)$ from being an accessible trajectory for the functional integral. We define $w^*(t;v)$ as the path that minimises Eq.~\eqref{eq:action_func_m}, $\left. \delta \mathcal{N}[w]/\delta w \right|_{w^*} = 0$, whence
\begin{equation} \label{eq:newt_map}
    m \ddot{w}^*(t) = \mu^2 (w^*(t) - \dot{x}_c(t) ) + \mu^2 m w^*(t)
\end{equation}
with $m = 1/{\rm Pe}$ and boundary condition $w^*(T)=v$. Equation~\eqref{eq:newt_map} is purposefully arranged to resemble the Newtonian dynamics of a particle of mass $m$ in an unstable, time-dependent harmonic potential $V(w^*,t) =  - [\mu^2(\dot{x}_c(t) - w^*)^2/2 + m \mu^2 {w^*}^2/2]$. Remarkably, the high-${\rm Pe}$ limit corresponds to the overdamped limit of Eq.~\eqref{eq:newt_map}, whereby $m \to 0$ and the potential term dominates. For $m \ll 1$, Eq.~\eqref{eq:newt_map} is solved by combining an exponential ansatz with the particular solution $w^*(t;v) = \dot{x}_c(t) + \mathcal{O}(m)$, whence
\begin{equation} \label{eq:m_exp_path}
    w^*(t) = \dot{x}_c(t) + (v-\dot{x}_c(T)) e^{ \sqrt{\frac{1+m}{m}} \mu (t-T)} + \mathcal{O}(m)~.
\end{equation} 
Noting the second functional derivative of $\mathcal{N}$ is independent of $w(t)$, we perform a change of variables $w(t) \to \delta w(t) + w^*(t;v)$ in the functional integral, Eq.~\eqref{eq:two_costs}, to rewrite Eq.~\eqref{eq:generic_vexpect} exactly as
\begin{equation} \label{eq:saddle_simpl}
    \mathbb{E}_w[ w(T) | \{x\}_0^T]
    =
    \frac{\int dv \ v \cdot e^{- \mathcal{N}[w^*(t;v);\{x\}_0^T]}] \mathbb{P}(v)}{\int dv \ e^{- \mathcal{N}[w^*(t;v);\{x\}_0^T]}] \mathbb{P}(v)}~.
\end{equation}
Substituting Eq.~\eqref{eq:m_exp_path} into Eq.~\eqref{eq:action_func_m} we thus have, to leading order in large ${\rm Pe}$, 
\begin{equation} \label{eq:saddle_func_eval}
    \mathcal{N}[w^*(t)]
    = \frac{\sqrt{\rm Pe}}{8} \left[ \left( \frac{v - \dot{x}_c(T)}{\sigma_w}\right)^2 + \mathcal{O}({\rm Pe}^{-\frac{1}{2}}) \right]~.
\end{equation}
which draws only on the potential term. Further substituting Eq.~\eqref{eq:saddle_func_eval} into Eq.~\eqref{eq:saddle_simpl} and performing all the resulting Gaussian integrals in closed form, we arrive at the following expression for the posterior expectation of the self-propulsion velocity at high ${\rm Pe}$,
\begin{equation}
    \mathbb{E}_w[ w(T) | \{x\}_0^T]  = \left( 1 - \frac{4}{\sqrt{\rm Pe}}\right) 
    \dot{x}_c(T)  + \mathcal{O}({\rm Pe}^{-1}) ~.
\end{equation} 
In other words, the prior distribution $\mathbb{P}[w]$ weakly biases our posterior estimation $\mathbb{E}_w[ w(T) | \{x\}_0^T]$ away from $\dot{x}_c(T)$ and towards the prior expectation $\mathbb{E}_w[ w(T)] =0$. Using Eq.~\eqref{eq:opt_w_colorless}, the high-Pe asymptotic average power output, having applied the optimal protocol, is thus given by
\begin{equation} \label{eq:upb_highPe_AOU}
    \mathbb{E}_{\xi,w}[\dot{W}_{\rm AOU}(F^*_{\rm ext})] = \frac{\sigma_w^2}{4} \left(1-\frac{8}{\sqrt{\rm Pe}} \right) + \mathcal{O}({\rm Pe}^{-1})~.
\end{equation}
\paragraph*{Conclusion ---} We have identified generic continuous feedback protocols achieving maximum average power extraction from active particles with a (zero-mean) hidden self-propulsion state. These optimal protocols can be written in closed form in the asymptotes ${\rm Pe} \ll 1$ and ${\rm Pe} \gg 1$, and provide upper bounds to the average extractable work by \emph{any} such protocol (cf.\ \cite{malgaretti2022}), e.g.\ Eqs.~\eqref{eq:RnT_lowPe_asym}, \eqref{eq:asym_lowPe-Aou} and \eqref{eq:upb_highPe_AOU}. These bounds are those against which the performance of autonomous active matter engines, which typically do not have access to the self-propulsion states of the individual constituent particles \cite{pietzonka2019autonomous,martinez2017colloidal}, should be compared. {Furthermore, our ``boundary-update" approach enables work extraction in experimental settings where real-time particle tracking is unfeasible, since only the detection of first-passage events is required for its implementation.}

The optimal protocol is generally non-Markovian. However, this difficulty can be circumvented at ${\rm Pe} \ll 1$ by embedding the dynamics in a higher dimensional phase space \cite{loos2019fokker}, e.g.\ via the auxiliary dynamics in Eq.~\eqref{eq:Qeq_resc}. Analogously to equilibrium information engines \cite{malgaretti2022,saha2022information}, the thermodynamic cost of operating the feedback control can be identified with the increase in the total entropy production rate upon expanding the phase space to include such auxiliary variables \cite{loos2020irreversibility}. In an idealised situation where the operating temperature of the measurement device is arbitrary, and can thus be chosen to be arbitrarily small, the associated dissipation is negligible \cite{saha2022information}. The unique utility of information engines operating on active particles arises from their non-vanishing efficiency even when the measurement device and the particle are coupled to the \textit{same} heat bath \cite{malgaretti2022}. Future work will characterise the efficiency of the optimal protocols in this case.

\begin{acknowledgments}
The authors thank Gunnar Pruessner, Farid Kaveh, Henry Alston and Zigan Zhen for useful discussions, and Yuning Chen for contributing to the schematic in Fig.~1. L.C.\ acknowledges support from the Francis Crick Institute, which receives its core funding from Cancer Research UK, the UK Medical Research Council, and the Wellcome Trust (FC001317). J.K.\ and C.R.\ acknowledge support from the Engineering and Physical Sciences Research Council (grant numbers 2620369 and 2478322, respectively).
\end{acknowledgments}

\bibliography{references}

\providecommand{\noopsort}[1]{}\providecommand{\singleletter}[1]{#1}%
\begin{thebibliography}{41}%
\makeatletter
\providecommand \@ifxundefined [1]{%
 \@ifx{#1\undefined}
}%
\providecommand \@ifnum [1]{%
 \ifnum #1\expandafter \@firstoftwo
 \else \expandafter \@secondoftwo
 \fi
}%
\providecommand \@ifx [1]{%
 \ifx #1\expandafter \@firstoftwo
 \else \expandafter \@secondoftwo
 \fi
}%
\providecommand \natexlab [1]{#1}%
\providecommand \enquote  [1]{``#1''}%
\providecommand \bibnamefont  [1]{#1}%
\providecommand \bibfnamefont [1]{#1}%
\providecommand \citenamefont [1]{#1}%
\providecommand \href@noop [0]{\@secondoftwo}%
\providecommand \href [0]{\begingroup \@sanitize@url \@href}%
\providecommand \@href[1]{\@@startlink{#1}\@@href}%
\providecommand \@@href[1]{\endgroup#1\@@endlink}%
\providecommand \@sanitize@url [0]{\catcode `\\12\catcode `\$12\catcode
  `\&12\catcode `\#12\catcode `\^12\catcode `\_12\catcode `\%12\relax}%
\providecommand \@@startlink[1]{}%
\providecommand \@@endlink[0]{}%
\providecommand \url  [0]{\begingroup\@sanitize@url \@url }%
\providecommand \@url [1]{\endgroup\@href {#1}{\urlprefix }}%
\providecommand \urlprefix  [0]{URL }%
\providecommand \Eprint [0]{\href }%
\providecommand \doibase [0]{https://doi.org/}%
\providecommand \selectlanguage [0]{\@gobble}%
\providecommand \bibinfo  [0]{\@secondoftwo}%
\providecommand \bibfield  [0]{\@secondoftwo}%
\providecommand \translation [1]{[#1]}%
\providecommand \BibitemOpen [0]{}%
\providecommand \bibitemStop [0]{}%
\providecommand \bibitemNoStop [0]{.\EOS\space}%
\providecommand \EOS [0]{\spacefactor3000\relax}%
\providecommand \BibitemShut  [1]{\csname bibitem#1\endcsname}%
\let\auto@bib@innerbib\@empty
\bibitem [{\citenamefont {Pietzonka}\ \emph {et~al.}(2019)\citenamefont
  {Pietzonka}, \citenamefont {Fodor}, \citenamefont {Lohrmann}, \citenamefont
  {Cates},\ and\ \citenamefont {Seifert}}]{pietzonka2019autonomous}%
  \BibitemOpen
  \bibfield  {author} {\bibinfo {author} {\bibfnamefont {P.}~\bibnamefont
  {Pietzonka}}, \bibinfo {author} {\bibfnamefont {{\'E}.}~\bibnamefont
  {Fodor}}, \bibinfo {author} {\bibfnamefont {C.}~\bibnamefont {Lohrmann}},
  \bibinfo {author} {\bibfnamefont {M.~E.}\ \bibnamefont {Cates}},\ and\
  \bibinfo {author} {\bibfnamefont {U.}~\bibnamefont {Seifert}},\ }\bibfield
  {title} {\bibinfo {title} {Autonomous engines driven by active matter:
  Energetics and design principles},\ }\href@noop {} {\bibfield  {journal}
  {\bibinfo  {journal} {Phys.\ Rev.\ X}\ }\textbf {\bibinfo {volume} {9}},\
  \bibinfo {pages} {041032} (\bibinfo {year} {2019})}\BibitemShut {NoStop}%
\bibitem [{\citenamefont {Mart{\'\i}nez}\ \emph {et~al.}(2017)\citenamefont
  {Mart{\'\i}nez}, \citenamefont {Rold{\'a}n}, \citenamefont {Dinis},\ and\
  \citenamefont {Rica}}]{martinez2017colloidal}%
  \BibitemOpen
  \bibfield  {author} {\bibinfo {author} {\bibfnamefont {I.~A.}\ \bibnamefont
  {Mart{\'\i}nez}}, \bibinfo {author} {\bibfnamefont {{\'E}.}~\bibnamefont
  {Rold{\'a}n}}, \bibinfo {author} {\bibfnamefont {L.}~\bibnamefont {Dinis}},\
  and\ \bibinfo {author} {\bibfnamefont {R.~A.}\ \bibnamefont {Rica}},\
  }\bibfield  {title} {\bibinfo {title} {Colloidal heat engines: A review},\
  }\href@noop {} {\bibfield  {journal} {\bibinfo  {journal} {Soft Matter}\
  }\textbf {\bibinfo {volume} {13}},\ \bibinfo {pages} {22} (\bibinfo {year}
  {2017})}\BibitemShut {NoStop}%
\bibitem [{\citenamefont {Mart{\'\i}nez}\ \emph {et~al.}(2016)\citenamefont
  {Mart{\'\i}nez}, \citenamefont {Rold{\'a}n}, \citenamefont {Dinis},
  \citenamefont {Petrov}, \citenamefont {Parrondo},\ and\ \citenamefont
  {Rica}}]{martinez2016brownian}%
  \BibitemOpen
  \bibfield  {author} {\bibinfo {author} {\bibfnamefont {I.~A.}\ \bibnamefont
  {Mart{\'\i}nez}}, \bibinfo {author} {\bibfnamefont {{\'E}.}~\bibnamefont
  {Rold{\'a}n}}, \bibinfo {author} {\bibfnamefont {L.}~\bibnamefont {Dinis}},
  \bibinfo {author} {\bibfnamefont {D.}~\bibnamefont {Petrov}}, \bibinfo
  {author} {\bibfnamefont {J.~M.}\ \bibnamefont {Parrondo}},\ and\ \bibinfo
  {author} {\bibfnamefont {R.~A.}\ \bibnamefont {Rica}},\ }\bibfield  {title}
  {\bibinfo {title} {Brownian {C}arnot engine},\ }\href@noop {} {\bibfield
  {journal} {\bibinfo  {journal} {Nat.\ Phys.}\ }\textbf {\bibinfo {volume}
  {12}},\ \bibinfo {pages} {67} (\bibinfo {year} {2016})}\BibitemShut {NoStop}%
\bibitem [{\citenamefont {Speck}(2022)}]{speck2022efficiency}%
  \BibitemOpen
  \bibfield  {author} {\bibinfo {author} {\bibfnamefont {T.}~\bibnamefont
  {Speck}},\ }\bibfield  {title} {\bibinfo {title} {Efficiency of isothermal
  active matter engines: Strong driving beats weak driving},\ }\href@noop {}
  {\bibfield  {journal} {\bibinfo  {journal} {Phys.\ Rev.\ E}\ }\textbf
  {\bibinfo {volume} {105}},\ \bibinfo {pages} {L012601} (\bibinfo {year}
  {2022})}\BibitemShut {NoStop}%
\bibitem [{\citenamefont {Pumm}\ \emph {et~al.}(2022)\citenamefont {Pumm},
  \citenamefont {Engelen}, \citenamefont {Kopperger}, \citenamefont {Isensee},
  \citenamefont {Vogt}, \citenamefont {Kozina}, \citenamefont {Kube},
  \citenamefont {Honemann}, \citenamefont {Bertosin}, \citenamefont {Langecker}
  \emph {et~al.}}]{pumm2022dna}%
  \BibitemOpen
  \bibfield  {author} {\bibinfo {author} {\bibfnamefont {A.-K.}\ \bibnamefont
  {Pumm}}, \bibinfo {author} {\bibfnamefont {W.}~\bibnamefont {Engelen}},
  \bibinfo {author} {\bibfnamefont {E.}~\bibnamefont {Kopperger}}, \bibinfo
  {author} {\bibfnamefont {J.}~\bibnamefont {Isensee}}, \bibinfo {author}
  {\bibfnamefont {M.}~\bibnamefont {Vogt}}, \bibinfo {author} {\bibfnamefont
  {V.}~\bibnamefont {Kozina}}, \bibinfo {author} {\bibfnamefont
  {M.}~\bibnamefont {Kube}}, \bibinfo {author} {\bibfnamefont {M.~N.}\
  \bibnamefont {Honemann}}, \bibinfo {author} {\bibfnamefont {E.}~\bibnamefont
  {Bertosin}}, \bibinfo {author} {\bibfnamefont {M.}~\bibnamefont {Langecker}},
  \emph {et~al.},\ }\bibfield  {title} {\bibinfo {title} {A {DNA} origami
  rotary ratchet motor},\ }\href@noop {} {\bibfield  {journal} {\bibinfo
  {journal} {Nature}\ }\textbf {\bibinfo {volume} {607}},\ \bibinfo {pages}
  {492} (\bibinfo {year} {2022})}\BibitemShut {NoStop}%
\bibitem [{\citenamefont {Malgaretti}\ and\ \citenamefont
  {Stark}(2022)}]{malgaretti2022}%
  \BibitemOpen
  \bibfield  {author} {\bibinfo {author} {\bibfnamefont {P.}~\bibnamefont
  {Malgaretti}}\ and\ \bibinfo {author} {\bibfnamefont {H.}~\bibnamefont
  {Stark}},\ }\bibfield  {title} {\bibinfo {title} {Szilard engines and
  information-based work extraction for active systems},\ }\href@noop {}
  {\bibfield  {journal} {\bibinfo  {journal} {Phys.\ Rev.\ Lett.}\ }\textbf
  {\bibinfo {volume} {129}},\ \bibinfo {pages} {228005} (\bibinfo {year}
  {2022})}\BibitemShut {NoStop}%
\bibitem [{\citenamefont {Di~Leonardo}\ \emph {et~al.}(2010)\citenamefont
  {Di~Leonardo}, \citenamefont {Angelani}, \citenamefont {Dell’Arciprete},
  \citenamefont {Ruocco}, \citenamefont {Iebba}, \citenamefont {Schippa},
  \citenamefont {Conte}, \citenamefont {Mecarini}, \citenamefont {De~Angelis},\
  and\ \citenamefont {Di~Fabrizio}}]{dileonardo2010bacterial}%
  \BibitemOpen
  \bibfield  {author} {\bibinfo {author} {\bibfnamefont {R.}~\bibnamefont
  {Di~Leonardo}}, \bibinfo {author} {\bibfnamefont {L.}~\bibnamefont
  {Angelani}}, \bibinfo {author} {\bibfnamefont {D.}~\bibnamefont
  {Dell’Arciprete}}, \bibinfo {author} {\bibfnamefont {G.}~\bibnamefont
  {Ruocco}}, \bibinfo {author} {\bibfnamefont {V.}~\bibnamefont {Iebba}},
  \bibinfo {author} {\bibfnamefont {S.}~\bibnamefont {Schippa}}, \bibinfo
  {author} {\bibfnamefont {M.~P.}\ \bibnamefont {Conte}}, \bibinfo {author}
  {\bibfnamefont {F.}~\bibnamefont {Mecarini}}, \bibinfo {author}
  {\bibfnamefont {F.}~\bibnamefont {De~Angelis}},\ and\ \bibinfo {author}
  {\bibfnamefont {E.}~\bibnamefont {Di~Fabrizio}},\ }\bibfield  {title}
  {\bibinfo {title} {Bacterial ratchet motors},\ }\href@noop {} {\bibfield
  {journal} {\bibinfo  {journal} {Proc.\ Natl.\ Acad.\ Sci.\ U.S.A.}\ }\textbf
  {\bibinfo {volume} {107}},\ \bibinfo {pages} {9541} (\bibinfo {year}
  {2010})}\BibitemShut {NoStop}%
\bibitem [{\citenamefont {Thampi}\ \emph {et~al.}(2016)\citenamefont {Thampi},
  \citenamefont {Doostmohammadi}, \citenamefont {Shendruk}, \citenamefont
  {Golestanian},\ and\ \citenamefont {Yeomans}}]{thampi2016active}%
  \BibitemOpen
  \bibfield  {author} {\bibinfo {author} {\bibfnamefont {S.~P.}\ \bibnamefont
  {Thampi}}, \bibinfo {author} {\bibfnamefont {A.}~\bibnamefont
  {Doostmohammadi}}, \bibinfo {author} {\bibfnamefont {T.~N.}\ \bibnamefont
  {Shendruk}}, \bibinfo {author} {\bibfnamefont {R.}~\bibnamefont
  {Golestanian}},\ and\ \bibinfo {author} {\bibfnamefont {J.~M.}\ \bibnamefont
  {Yeomans}},\ }\bibfield  {title} {\bibinfo {title} {Active micromachines:
  Microfluidics powered by mesoscale turbulence},\ }\href@noop {} {\bibfield
  {journal} {\bibinfo  {journal} {Sci.\ Adv.}\ }\textbf {\bibinfo {volume}
  {2}},\ \bibinfo {pages} {e1501854} (\bibinfo {year} {2016})}\BibitemShut
  {NoStop}%
\bibitem [{\citenamefont {P{\"o}nisch}\ and\ \citenamefont
  {Zaburdaev}(2022)}]{ponisch2022pili}%
  \BibitemOpen
  \bibfield  {author} {\bibinfo {author} {\bibfnamefont {W.}~\bibnamefont
  {P{\"o}nisch}}\ and\ \bibinfo {author} {\bibfnamefont {V.}~\bibnamefont
  {Zaburdaev}},\ }\bibfield  {title} {\bibinfo {title} {A pili-driven bacterial
  turbine},\ }\href@noop {} {\bibfield  {journal} {\bibinfo  {journal} {Front.\
  Phys.}\ }\textbf {\bibinfo {volume} {10}},\ \bibinfo {pages} {875687}
  (\bibinfo {year} {2022})}\BibitemShut {NoStop}%
\bibitem [{\citenamefont {Zhen}\ and\ \citenamefont
  {Pruessner}(2022)}]{zhen2022optimal}%
  \BibitemOpen
  \bibfield  {author} {\bibinfo {author} {\bibfnamefont {Z.}~\bibnamefont
  {Zhen}}\ and\ \bibinfo {author} {\bibfnamefont {G.}~\bibnamefont
  {Pruessner}},\ }\bibfield  {title} {\bibinfo {title} {Optimal ratchet
  potentials for run-and-tumble particles},\ }\href@noop {} {\bibfield
  {journal} {\bibinfo  {journal} {arXiv:2204.04070}\ } (\bibinfo {year}
  {2022})}\BibitemShut {NoStop}%
\bibitem [{\citenamefont {Berger}\ \emph {et~al.}(2009)\citenamefont {Berger},
  \citenamefont {Schmiedl},\ and\ \citenamefont {Seifert}}]{berger2009optimal}%
  \BibitemOpen
  \bibfield  {author} {\bibinfo {author} {\bibfnamefont {F.}~\bibnamefont
  {Berger}}, \bibinfo {author} {\bibfnamefont {T.}~\bibnamefont {Schmiedl}},\
  and\ \bibinfo {author} {\bibfnamefont {U.}~\bibnamefont {Seifert}},\
  }\bibfield  {title} {\bibinfo {title} {Optimal potentials for temperature
  ratchets},\ }\href@noop {} {\bibfield  {journal} {\bibinfo  {journal} {Phys.\
  Rev.\ E}\ }\textbf {\bibinfo {volume} {79}},\ \bibinfo {pages} {031118}
  (\bibinfo {year} {2009})}\BibitemShut {NoStop}%
\bibitem [{\citenamefont {Esposito}(2012)}]{esposito2012stochastic}%
  \BibitemOpen
  \bibfield  {author} {\bibinfo {author} {\bibfnamefont {M.}~\bibnamefont
  {Esposito}},\ }\bibfield  {title} {\bibinfo {title} {Stochastic
  thermodynamics under coarse graining},\ }\href@noop {} {\bibfield  {journal}
  {\bibinfo  {journal} {Phys.\ Rev.\ E}\ }\textbf {\bibinfo {volume} {85}},\
  \bibinfo {pages} {041125} (\bibinfo {year} {2012})}\BibitemShut {NoStop}%
\bibitem [{\citenamefont {Onsager}\ and\ \citenamefont
  {Machlup}(1953)}]{onsager1953fluctuations}%
  \BibitemOpen
  \bibfield  {author} {\bibinfo {author} {\bibfnamefont {L.}~\bibnamefont
  {Onsager}}\ and\ \bibinfo {author} {\bibfnamefont {S.}~\bibnamefont
  {Machlup}},\ }\bibfield  {title} {\bibinfo {title} {Fluctuations and
  irreversible processes},\ }\href@noop {} {\bibfield  {journal} {\bibinfo
  {journal} {Phys.\ Rev.}\ }\textbf {\bibinfo {volume} {91}},\ \bibinfo {pages}
  {1505} (\bibinfo {year} {1953})}\BibitemShut {NoStop}%
\bibitem [{\citenamefont {Garcia-Millan}\ and\ \citenamefont
  {Pruessner}(2021)}]{garcia2021run}%
  \BibitemOpen
  \bibfield  {author} {\bibinfo {author} {\bibfnamefont {R.}~\bibnamefont
  {Garcia-Millan}}\ and\ \bibinfo {author} {\bibfnamefont {G.}~\bibnamefont
  {Pruessner}},\ }\bibfield  {title} {\bibinfo {title} {Run-and-tumble motion
  in a harmonic potential: {F}ield theory and entropy production},\ }\href@noop
  {} {\bibfield  {journal} {\bibinfo  {journal} {J.\ Stat.\ Mech.}\ }\textbf
  {\bibinfo {volume} {2021}},\ \bibinfo {pages} {063203} (\bibinfo {year}
  {2021})}\BibitemShut {NoStop}%
\bibitem [{\citenamefont {Martin}\ \emph {et~al.}(2021)\citenamefont {Martin},
  \citenamefont {O'Byrne}, \citenamefont {Cates}, \citenamefont {Fodor},
  \citenamefont {Nardini}, \citenamefont {Tailleur},\ and\ \citenamefont {van
  Wijland}}]{martin2021statistical}%
  \BibitemOpen
  \bibfield  {author} {\bibinfo {author} {\bibfnamefont {D.}~\bibnamefont
  {Martin}}, \bibinfo {author} {\bibfnamefont {J.}~\bibnamefont {O'Byrne}},
  \bibinfo {author} {\bibfnamefont {M.~E.}\ \bibnamefont {Cates}}, \bibinfo
  {author} {\bibfnamefont {{\'E}.}~\bibnamefont {Fodor}}, \bibinfo {author}
  {\bibfnamefont {C.}~\bibnamefont {Nardini}}, \bibinfo {author} {\bibfnamefont
  {J.}~\bibnamefont {Tailleur}},\ and\ \bibinfo {author} {\bibfnamefont
  {F.}~\bibnamefont {van Wijland}},\ }\bibfield  {title} {\bibinfo {title}
  {Statistical mechanics of active {O}rnstein-{U}hlenbeck particles},\
  }\href@noop {} {\bibfield  {journal} {\bibinfo  {journal} {Phys.\ Rev.\ E}\
  }\textbf {\bibinfo {volume} {103}},\ \bibinfo {pages} {032607} (\bibinfo
  {year} {2021})}\BibitemShut {NoStop}%
\bibitem [{\citenamefont {Fodor}\ \emph {et~al.}(2022)\citenamefont {Fodor},
  \citenamefont {Jack},\ and\ \citenamefont
  {Cates}}]{fodor2022irreversibility}%
  \BibitemOpen
  \bibfield  {author} {\bibinfo {author} {\bibfnamefont {{\'E}.}~\bibnamefont
  {Fodor}}, \bibinfo {author} {\bibfnamefont {R.~L.}\ \bibnamefont {Jack}},\
  and\ \bibinfo {author} {\bibfnamefont {M.~E.}\ \bibnamefont {Cates}},\
  }\bibfield  {title} {\bibinfo {title} {Irreversibility and biased ensembles
  in active matter: Insights from stochastic thermodynamics},\ }\href@noop {}
  {\bibfield  {journal} {\bibinfo  {journal} {Annu.\ Rev.\ Condens.\ Matter
  Phys.}\ }\textbf {\bibinfo {volume} {13}},\ \bibinfo {pages} {215} (\bibinfo
  {year} {2022})}\BibitemShut {NoStop}%
\bibitem [{\citenamefont {Markovich}\ \emph {et~al.}(2021)\citenamefont
  {Markovich}, \citenamefont {Fodor}, \citenamefont {Tjhung},\ and\
  \citenamefont {Cates}}]{markovich2021thermodynamics}%
  \BibitemOpen
  \bibfield  {author} {\bibinfo {author} {\bibfnamefont {T.}~\bibnamefont
  {Markovich}}, \bibinfo {author} {\bibfnamefont {{\'E}.}~\bibnamefont
  {Fodor}}, \bibinfo {author} {\bibfnamefont {E.}~\bibnamefont {Tjhung}},\ and\
  \bibinfo {author} {\bibfnamefont {M.~E.}\ \bibnamefont {Cates}},\ }\bibfield
  {title} {\bibinfo {title} {Thermodynamics of active field theories: Energetic
  cost of coupling to reservoirs},\ }\href@noop {} {\bibfield  {journal}
  {\bibinfo  {journal} {Phys.\ Rev.\ X}\ }\textbf {\bibinfo {volume} {11}},\
  \bibinfo {pages} {021057} (\bibinfo {year} {2021})}\BibitemShut {NoStop}%
\bibitem [{\citenamefont {Kullback}\ and\ \citenamefont
  {Leibler}(1951)}]{kullback1951information}%
  \BibitemOpen
  \bibfield  {author} {\bibinfo {author} {\bibfnamefont {S.}~\bibnamefont
  {Kullback}}\ and\ \bibinfo {author} {\bibfnamefont {R.~A.}\ \bibnamefont
  {Leibler}},\ }\bibfield  {title} {\bibinfo {title} {On information and
  sufficiency},\ }\href@noop {} {\bibfield  {journal} {\bibinfo  {journal}
  {Ann.\ Math.\ Stat.}\ }\textbf {\bibinfo {volume} {22}},\ \bibinfo {pages}
  {79} (\bibinfo {year} {1951})}\BibitemShut {NoStop}%
\bibitem [{\citenamefont {Gaspard}(2004)}]{gaspard2004time}%
  \BibitemOpen
  \bibfield  {author} {\bibinfo {author} {\bibfnamefont {P.}~\bibnamefont
  {Gaspard}},\ }\bibfield  {title} {\bibinfo {title} {Time-reversed dynamical
  entropy and irreversibility in {M}arkovian random processes},\ }\href@noop {}
  {\bibfield  {journal} {\bibinfo  {journal} {J.\ Stat.\ Phys.}\ }\textbf
  {\bibinfo {volume} {117}},\ \bibinfo {pages} {599} (\bibinfo {year}
  {2004})}\BibitemShut {NoStop}%
\bibitem [{\citenamefont {Rold{\'a}n}\ \emph {et~al.}(2015)\citenamefont
  {Rold{\'a}n}, \citenamefont {Neri}, \citenamefont {D{\"o}rpinghaus},
  \citenamefont {Meyr},\ and\ \citenamefont
  {J{\"u}licher}}]{roldan2015decision}%
  \BibitemOpen
  \bibfield  {author} {\bibinfo {author} {\bibfnamefont {{\'E}.}~\bibnamefont
  {Rold{\'a}n}}, \bibinfo {author} {\bibfnamefont {I.}~\bibnamefont {Neri}},
  \bibinfo {author} {\bibfnamefont {M.}~\bibnamefont {D{\"o}rpinghaus}},
  \bibinfo {author} {\bibfnamefont {H.}~\bibnamefont {Meyr}},\ and\ \bibinfo
  {author} {\bibfnamefont {F.}~\bibnamefont {J{\"u}licher}},\ }\bibfield
  {title} {\bibinfo {title} {Decision making in the arrow of time},\
  }\href@noop {} {\bibfield  {journal} {\bibinfo  {journal} {Phys.\ Rev.\
  Lett.}\ }\textbf {\bibinfo {volume} {115}},\ \bibinfo {pages} {250602}
  (\bibinfo {year} {2015})}\BibitemShut {NoStop}%
\bibitem [{\citenamefont {Cocconi}\ \emph {et~al.}(2022)\citenamefont
  {Cocconi}, \citenamefont {Salbreux},\ and\ \citenamefont
  {Pruessner}}]{cocconi2022scaling}%
  \BibitemOpen
  \bibfield  {author} {\bibinfo {author} {\bibfnamefont {L.}~\bibnamefont
  {Cocconi}}, \bibinfo {author} {\bibfnamefont {G.}~\bibnamefont {Salbreux}},\
  and\ \bibinfo {author} {\bibfnamefont {G.}~\bibnamefont {Pruessner}},\
  }\bibfield  {title} {\bibinfo {title} {Scaling of entropy production under
  coarse graining in active disordered media},\ }\href@noop {} {\bibfield
  {journal} {\bibinfo  {journal} {Phys.\ Rev.\ E}\ }\textbf {\bibinfo {volume}
  {105}},\ \bibinfo {pages} {L042601} (\bibinfo {year} {2022})}\BibitemShut
  {NoStop}%
\bibitem [{\citenamefont {Yu}\ \emph {et~al.}(2021)\citenamefont {Yu},
  \citenamefont {Zhang},\ and\ \citenamefont {Tu}}]{yu2021inverse}%
  \BibitemOpen
  \bibfield  {author} {\bibinfo {author} {\bibfnamefont {Q.}~\bibnamefont
  {Yu}}, \bibinfo {author} {\bibfnamefont {D.}~\bibnamefont {Zhang}},\ and\
  \bibinfo {author} {\bibfnamefont {Y.}~\bibnamefont {Tu}},\ }\bibfield
  {title} {\bibinfo {title} {Inverse power law scaling of energy dissipation
  rate in nonequilibrium reaction networks},\ }\href@noop {} {\bibfield
  {journal} {\bibinfo  {journal} {Phys.\ Rev.\ Lett.}\ }\textbf {\bibinfo
  {volume} {126}},\ \bibinfo {pages} {080601} (\bibinfo {year}
  {2021})}\BibitemShut {NoStop}%
\bibitem [{\citenamefont {Ro}\ \emph {et~al.}(2022)\citenamefont {Ro},
  \citenamefont {Guo}, \citenamefont {Shih}, \citenamefont {Phan},
  \citenamefont {Austin}, \citenamefont {Levine}, \citenamefont {Chaikin},\
  and\ \citenamefont {Martiniani}}]{PhysRevLett.129.220601}%
  \BibitemOpen
  \bibfield  {author} {\bibinfo {author} {\bibfnamefont {S.}~\bibnamefont
  {Ro}}, \bibinfo {author} {\bibfnamefont {B.}~\bibnamefont {Guo}}, \bibinfo
  {author} {\bibfnamefont {A.}~\bibnamefont {Shih}}, \bibinfo {author}
  {\bibfnamefont {T.~V.}\ \bibnamefont {Phan}}, \bibinfo {author}
  {\bibfnamefont {R.~H.}\ \bibnamefont {Austin}}, \bibinfo {author}
  {\bibfnamefont {D.}~\bibnamefont {Levine}}, \bibinfo {author} {\bibfnamefont
  {P.~M.}\ \bibnamefont {Chaikin}},\ and\ \bibinfo {author} {\bibfnamefont
  {S.}~\bibnamefont {Martiniani}},\ }\bibfield  {title} {\bibinfo {title}
  {Model-free measurement of local entropy production and extractable work in
  active matter},\ }\href@noop {} {\bibfield  {journal} {\bibinfo  {journal}
  {Phys.\ Rev.\ Lett.}\ }\textbf {\bibinfo {volume} {129}},\ \bibinfo {pages}
  {220601} (\bibinfo {year} {2022})}\BibitemShut {NoStop}%
\bibitem [{\citenamefont {Leff}\ and\ \citenamefont
  {Rex}(2002)}]{leff2002maxwell}%
  \BibitemOpen
  \bibfield  {author} {\bibinfo {author} {\bibfnamefont {H.}~\bibnamefont
  {Leff}}\ and\ \bibinfo {author} {\bibfnamefont {A.~F.}\ \bibnamefont {Rex}},\
  }\href@noop {} {\emph {\bibinfo {title} {Maxwell's Demon 2 Entropy, Classical
  and Quantum Information, Computing}}}\ (\bibinfo  {publisher} {CRC Press},\
  \bibinfo {year} {2002})\BibitemShut {NoStop}%
\bibitem [{\citenamefont {Ouldridge}\ \emph {et~al.}(2018)\citenamefont
  {Ouldridge}, \citenamefont {Brittain},\ and\ \citenamefont
  {Wolde}}]{ouldridge_power_2018}%
  \BibitemOpen
  \bibfield  {author} {\bibinfo {author} {\bibfnamefont {T.~E.}\ \bibnamefont
  {Ouldridge}}, \bibinfo {author} {\bibfnamefont {R.~A.}\ \bibnamefont
  {Brittain}},\ and\ \bibinfo {author} {\bibfnamefont {P.~R.~t.}\ \bibnamefont
  {Wolde}},\ }\bibfield  {title} {\bibinfo {title} {The power of being
  explicit: {D}emystifying work, heat, and free energy in the physics of
  computation},\ }in\ \href@noop {} {\emph {\bibinfo {booktitle} {The
  Energetics of Computing in Life and Machines}}},\ \bibinfo {editor} {edited
  by\ \bibinfo {editor} {\bibfnamefont {D.~H.}\ \bibnamefont {Wolpert}}}\
  (\bibinfo  {publisher} {SFI Press},\ \bibinfo {year} {2018})\BibitemShut
  {NoStop}%
\bibitem [{\citenamefont {Saha}\ \emph {et~al.}(2023)\citenamefont {Saha},
  \citenamefont {Ehrich}, \citenamefont {Gavrilov}, \citenamefont {Still},
  \citenamefont {Sivak},\ and\ \citenamefont
  {Bechhoefer}}]{saha2022information}%
  \BibitemOpen
  \bibfield  {author} {\bibinfo {author} {\bibfnamefont {T.~K.}\ \bibnamefont
  {Saha}}, \bibinfo {author} {\bibfnamefont {J.}~\bibnamefont {Ehrich}},
  \bibinfo {author} {\bibfnamefont {M.}~\bibnamefont {Gavrilov}}, \bibinfo
  {author} {\bibfnamefont {S.}~\bibnamefont {Still}}, \bibinfo {author}
  {\bibfnamefont {D.~A.}\ \bibnamefont {Sivak}},\ and\ \bibinfo {author}
  {\bibfnamefont {J.}~\bibnamefont {Bechhoefer}},\ }\bibfield  {title}
  {\bibinfo {title} {Information engine in a nonequilibrium bath},\ }\href@noop
  {} {\bibfield  {journal} {\bibinfo  {journal} {Phys.\ Rev.\ Lett.}\ }\textbf
  {\bibinfo {volume} {131}},\ \bibinfo {pages} {057101} (\bibinfo {year}
  {2023})}\BibitemShut {NoStop}%
\bibitem [{\citenamefont {Sandoval}\ \emph {et~al.}(2016)\citenamefont
  {Sandoval}, \citenamefont {Velasco},\ and\ \citenamefont
  {Jim{\'e}nez-Aquino}}]{sandoval2016magnetic}%
  \BibitemOpen
  \bibfield  {author} {\bibinfo {author} {\bibfnamefont {M.}~\bibnamefont
  {Sandoval}}, \bibinfo {author} {\bibfnamefont {R.}~\bibnamefont {Velasco}},\
  and\ \bibinfo {author} {\bibfnamefont {J.}~\bibnamefont
  {Jim{\'e}nez-Aquino}},\ }\bibfield  {title} {\bibinfo {title} {Magnetic field
  effect on charged brownian swimmers},\ }\href@noop {} {\bibfield  {journal}
  {\bibinfo  {journal} {Physica A}\ }\textbf {\bibinfo {volume} {442}},\
  \bibinfo {pages} {321} (\bibinfo {year} {2016})}\BibitemShut {NoStop}%
\bibitem [{\citenamefont {Roberts}\ and\ \citenamefont
  {Zhen}(2023)}]{roberts2023run}%
  \BibitemOpen
  \bibfield  {author} {\bibinfo {author} {\bibfnamefont {C.}~\bibnamefont
  {Roberts}}\ and\ \bibinfo {author} {\bibfnamefont {Z.}~\bibnamefont {Zhen}},\
  }\bibfield  {title} {\bibinfo {title} {Run-and-tumble motion in a linear
  ratchet potential: Analytic solution, power extraction, and first-passage
  properties},\ }\href@noop {} {\bibfield  {journal} {\bibinfo  {journal}
  {Phys.\ Rev.\ E}\ }\textbf {\bibinfo {volume} {108}},\ \bibinfo {pages}
  {014139} (\bibinfo {year} {2023})}\BibitemShut {NoStop}%
\bibitem [{\citenamefont {Seifert}(2012)}]{seifert2012stochastic}%
  \BibitemOpen
  \bibfield  {author} {\bibinfo {author} {\bibfnamefont {U.}~\bibnamefont
  {Seifert}},\ }\bibfield  {title} {\bibinfo {title} {Stochastic
  thermodynamics, fluctuation theorems and molecular machines},\ }\href@noop {}
  {\bibfield  {journal} {\bibinfo  {journal} {Rep.\ Prog.\ Phys.}\ }\textbf
  {\bibinfo {volume} {75}},\ \bibinfo {pages} {126001} (\bibinfo {year}
  {2012})}\BibitemShut {NoStop}%
\bibitem [{\citenamefont {Cocconi}\ \emph {et~al.}(2020)\citenamefont
  {Cocconi}, \citenamefont {Garcia-Millan}, \citenamefont {Zhen}, \citenamefont
  {Buturca},\ and\ \citenamefont {Pruessner}}]{cocconi2020entropy}%
  \BibitemOpen
  \bibfield  {author} {\bibinfo {author} {\bibfnamefont {L.}~\bibnamefont
  {Cocconi}}, \bibinfo {author} {\bibfnamefont {R.}~\bibnamefont
  {Garcia-Millan}}, \bibinfo {author} {\bibfnamefont {Z.}~\bibnamefont {Zhen}},
  \bibinfo {author} {\bibfnamefont {B.}~\bibnamefont {Buturca}},\ and\ \bibinfo
  {author} {\bibfnamefont {G.}~\bibnamefont {Pruessner}},\ }\bibfield  {title}
  {\bibinfo {title} {Entropy production in exactly solvable systems},\
  }\href@noop {} {\bibfield  {journal} {\bibinfo  {journal} {Entropy}\ }\textbf
  {\bibinfo {volume} {22}},\ \bibinfo {pages} {1252} (\bibinfo {year}
  {2020})}\BibitemShut {NoStop}%
\bibitem [{\citenamefont {Le~Bellac}(1991)}]{lebellac1991}%
  \BibitemOpen
  \bibfield  {author} {\bibinfo {author} {\bibfnamefont {M.}~\bibnamefont
  {Le~Bellac}},\ }\href@noop {} {\emph {\bibinfo {title} {Quantum and
  statistical field theory}}}\ (\bibinfo  {publisher} {Clarendon Press, New
  York},\ \bibinfo {year} {1991})\BibitemShut {NoStop}%
\bibitem [{\citenamefont {Bothe}\ \emph {et~al.}(2023)\citenamefont {Bothe},
  \citenamefont {Cocconi}, \citenamefont {Zhen},\ and\ \citenamefont
  {Pruessner}}]{bothe2022particle}%
  \BibitemOpen
  \bibfield  {author} {\bibinfo {author} {\bibfnamefont {M.}~\bibnamefont
  {Bothe}}, \bibinfo {author} {\bibfnamefont {L.}~\bibnamefont {Cocconi}},
  \bibinfo {author} {\bibfnamefont {Z.}~\bibnamefont {Zhen}},\ and\ \bibinfo
  {author} {\bibfnamefont {G.}~\bibnamefont {Pruessner}},\ }\bibfield  {title}
  {\bibinfo {title} {Particle entity in the {D}oi-{P}eliti and response field
  formalisms},\ }\href@noop {} {\bibfield  {journal} {\bibinfo  {journal} {J.\
  Phys.\ A}\ }\textbf {\bibinfo {volume} {56}},\ \bibinfo {pages} {175002}
  (\bibinfo {year} {2023})}\BibitemShut {NoStop}%
\bibitem [{\citenamefont {Caprini}\ \emph {et~al.}(2022)\citenamefont
  {Caprini}, \citenamefont {Sprenger}, \citenamefont {L{\"o}wen},\ and\
  \citenamefont {Wittmann}}]{caprini2022parental}%
  \BibitemOpen
  \bibfield  {author} {\bibinfo {author} {\bibfnamefont {L.}~\bibnamefont
  {Caprini}}, \bibinfo {author} {\bibfnamefont {A.~R.}\ \bibnamefont
  {Sprenger}}, \bibinfo {author} {\bibfnamefont {H.}~\bibnamefont
  {L{\"o}wen}},\ and\ \bibinfo {author} {\bibfnamefont {R.}~\bibnamefont
  {Wittmann}},\ }\bibfield  {title} {\bibinfo {title} {The parental active
  model: A unifying stochastic description of self-propulsion},\ }\href@noop {}
  {\bibfield  {journal} {\bibinfo  {journal} {J.\ Chem.\ Phys.}\ }\textbf
  {\bibinfo {volume} {156}},\ \bibinfo {pages} {071102} (\bibinfo {year}
  {2022})}\BibitemShut {NoStop}%
\bibitem [{\citenamefont {Loos}\ and\ \citenamefont
  {Klapp}(2019)}]{loos2019fokker}%
  \BibitemOpen
  \bibfield  {author} {\bibinfo {author} {\bibfnamefont {S.~A.}\ \bibnamefont
  {Loos}}\ and\ \bibinfo {author} {\bibfnamefont {S.~H.}\ \bibnamefont
  {Klapp}},\ }\bibfield  {title} {\bibinfo {title} {{F}okker-{P}lanck equations
  for time-delayed systems via {M}arkovian embedding},\ }\href@noop {}
  {\bibfield  {journal} {\bibinfo  {journal} {J.\ Stat.\ Phys.}\ }\textbf
  {\bibinfo {volume} {177}},\ \bibinfo {pages} {95} (\bibinfo {year}
  {2019})}\BibitemShut {NoStop}%
\bibitem [{\citenamefont {Loos}\ and\ \citenamefont
  {Klapp}(2020)}]{loos2020irreversibility}%
  \BibitemOpen
  \bibfield  {author} {\bibinfo {author} {\bibfnamefont {S.~A.}\ \bibnamefont
  {Loos}}\ and\ \bibinfo {author} {\bibfnamefont {S.~H.}\ \bibnamefont
  {Klapp}},\ }\bibfield  {title} {\bibinfo {title} {Irreversibility, heat and
  information flows induced by non-reciprocal interactions},\ }\href@noop {}
  {\bibfield  {journal} {\bibinfo  {journal} {New J.\ Phys.}\ }\textbf
  {\bibinfo {volume} {22}},\ \bibinfo {pages} {123051} (\bibinfo {year}
  {2020})}\BibitemShut {NoStop}%
\bibitem [{\citenamefont {Medeiros}\ and\ \citenamefont
  {Queir{\'o}s}(2021)}]{medeiros2021effective}%
  \BibitemOpen
  \bibfield  {author} {\bibinfo {author} {\bibfnamefont {J.~R.}\ \bibnamefont
  {Medeiros}}\ and\ \bibinfo {author} {\bibfnamefont {S.~M.~D.}\ \bibnamefont
  {Queir{\'o}s}},\ }\bibfield  {title} {\bibinfo {title} {Effective
  temperatures for single particle system under dichotomous noise},\
  }\href@noop {} {\bibfield  {journal} {\bibinfo  {journal} {J.\ Stat.\ Mech.}\
  }\textbf {\bibinfo {volume} {2021}},\ \bibinfo {pages} {063205} (\bibinfo
  {year} {2021})}\BibitemShut {NoStop}%
\bibitem [{\citenamefont {Celani}\ \emph {et~al.}(2012)\citenamefont {Celani},
  \citenamefont {Bo}, \citenamefont {Eichhorn},\ and\ \citenamefont
  {Aurell}}]{celani2012anomalous}%
  \BibitemOpen
  \bibfield  {author} {\bibinfo {author} {\bibfnamefont {A.}~\bibnamefont
  {Celani}}, \bibinfo {author} {\bibfnamefont {S.}~\bibnamefont {Bo}}, \bibinfo
  {author} {\bibfnamefont {R.}~\bibnamefont {Eichhorn}},\ and\ \bibinfo
  {author} {\bibfnamefont {E.}~\bibnamefont {Aurell}},\ }\bibfield  {title}
  {\bibinfo {title} {Anomalous thermodynamics at the microscale},\ }\href@noop
  {} {\bibfield  {journal} {\bibinfo  {journal} {Phys.\ Rev.\ Lett.}\ }\textbf
  {\bibinfo {volume} {109}},\ \bibinfo {pages} {260603} (\bibinfo {year}
  {2012})}\BibitemShut {NoStop}%
\bibitem [{\citenamefont {Gardiner}\ \emph {et~al.}(1985)\citenamefont
  {Gardiner} \emph {et~al.}}]{gardiner1985handbook}%
  \BibitemOpen
  \bibfield  {author} {\bibinfo {author} {\bibfnamefont {C.~W.}\ \bibnamefont
  {Gardiner}} \emph {et~al.},\ }\href@noop {} {\emph {\bibinfo {title}
  {Handbook of stochastic methods}}},\ Vol.~\bibinfo {volume} {3}\ (\bibinfo
  {publisher} {Springer, Berlin},\ \bibinfo {year} {1985})\BibitemShut
  {NoStop}%
\bibitem [{\citenamefont {Andrews}\ \emph {et~al.}(2013)\citenamefont
  {Andrews}, \citenamefont {Billingsley}, \citenamefont {Fiasconaro},
  \citenamefont {Frieden}, \citenamefont {Read}, \citenamefont {Shanks},\ and\
  \citenamefont {Treitel}}]{andrews2013picture}%
  \BibitemOpen
  \bibfield  {author} {\bibinfo {author} {\bibfnamefont {H.~C.}\ \bibnamefont
  {Andrews}}, \bibinfo {author} {\bibfnamefont {F.}~\bibnamefont
  {Billingsley}}, \bibinfo {author} {\bibfnamefont {J.}~\bibnamefont
  {Fiasconaro}}, \bibinfo {author} {\bibfnamefont {B.}~\bibnamefont {Frieden}},
  \bibinfo {author} {\bibfnamefont {R.}~\bibnamefont {Read}}, \bibinfo {author}
  {\bibfnamefont {J.}~\bibnamefont {Shanks}},\ and\ \bibinfo {author}
  {\bibfnamefont {S.}~\bibnamefont {Treitel}},\ }\href@noop {} {\emph {\bibinfo
  {title} {Picture processing and digital filtering}}},\ Vol.~\bibinfo {volume}
  {6}\ (\bibinfo  {publisher} {Springer Science \& Business Media, New York},\
  \bibinfo {year} {2013})\BibitemShut {NoStop}%
\bibitem [{\citenamefont {Malakar}\ \emph {et~al.}(2018)\citenamefont
  {Malakar}, \citenamefont {Jemseena}, \citenamefont {Kundu}, \citenamefont
  {Kumar}, \citenamefont {Sabhapandit}, \citenamefont {Majumdar}, \citenamefont
  {Redner},\ and\ \citenamefont {Dhar}}]{malakar2018steady}%
  \BibitemOpen
  \bibfield  {author} {\bibinfo {author} {\bibfnamefont {K.}~\bibnamefont
  {Malakar}}, \bibinfo {author} {\bibfnamefont {V.}~\bibnamefont {Jemseena}},
  \bibinfo {author} {\bibfnamefont {A.}~\bibnamefont {Kundu}}, \bibinfo
  {author} {\bibfnamefont {K.~V.}\ \bibnamefont {Kumar}}, \bibinfo {author}
  {\bibfnamefont {S.}~\bibnamefont {Sabhapandit}}, \bibinfo {author}
  {\bibfnamefont {S.~N.}\ \bibnamefont {Majumdar}}, \bibinfo {author}
  {\bibfnamefont {S.}~\bibnamefont {Redner}},\ and\ \bibinfo {author}
  {\bibfnamefont {A.}~\bibnamefont {Dhar}},\ }\bibfield  {title} {\bibinfo
  {title} {Steady state, relaxation and first-passage properties of a
  run-and-tumble particle in one-dimension},\ }\href@noop {} {\bibfield
  {journal} {\bibinfo  {journal} {J.\ Stat.\ Mech.}\ }\textbf {\bibinfo
  {volume} {2018}},\ \bibinfo {pages} {043215} (\bibinfo {year}
  {2018})}\BibitemShut {NoStop}%
\bibitem [{\citenamefont {Van~Kampen}(1992)}]{van1992stochastic}%
  \BibitemOpen
  \bibfield  {author} {\bibinfo {author} {\bibfnamefont {N.~G.}\ \bibnamefont
  {Van~Kampen}},\ }\href@noop {} {\emph {\bibinfo {title} {Stochastic processes
  in physics and chemistry}}},\ Vol.~\bibinfo {volume} {1}\ (\bibinfo
  {publisher} {Elsevier, New York},\ \bibinfo {year} {1992})\BibitemShut
  {NoStop}%
\end{thebibliography}%




\onecolumngrid

\section*{Supplementary Material}

\renewcommand{\theequation}{S\arabic{equation}}
\setcounter{equation}{0}

\renewcommand{\thefigure}{S\arabic{figure}}
\setcounter{figure}{0}

\renewcommand{\thesection}{S\Roman{section}}
\setcounter{section}{0}

\section{Time-reversal symmetry of active motion with hidden self-propulsion} \label{sr:I}

Consider a motile active particle with dynamics governed by the Langevin equation $\dot{x}(t) = w(t) + \sqrt{2D_x} \eta(t)$, where $\eta(t)$ is a unit covariance white noise, $D_x$ the spatial diffusivity and $w(t)$ the fluctuating self-propulsion velocity which we take to be inaccessible to direct observation, i.e.\ to constitute a hidden state. In this supplementary section, we show that, when $w(t)$ is a \emph{parity-time} (PT) symmetric stochastic process, the ensuing $x$-dynamics are time-reversal symmetric. This result applies to the particular cases of zero-mean-drift RnT and AOU dynamics discussed in the main text, where $w(t)$ is given by a telegraph process and an Ornstein-Uhlenbeck process respectively, among others \cite{caprini2022parental}. Both of these are separately parity \emph{and} time (and thus PT) symmetric. As a measure of time-reversal symmetry, we take the Kullback-Leibler divergence \cite{kullback1951information} of the ensemble of forward paths and their time-reversed counterparts, which also defines the average total informational entropy produced along a fluctuating trajectory \cite{fodor2022irreversibility,markovich2021thermodynamics},
\begin{equation} \label{eq:si_inf}
    \Delta S[\{x\}_0^T] = \int \mathcal{D}x \ \mathbb{P}[\{x\}_0^T] \ln \left(  \frac{\mathbb{P}[\{x\}_0^T]}{\mathbb{P}[\{x^R\}_0^T} \right)~.
\end{equation}
Here, $\mathbb{P}[\{x\}_0^T]$ denotes the probability density of a particular positional path and $\dot{x}^R(t) = -\dot{x}(T-t)$ denotes the time-reversed velocity, with the negative sign originating from odd parity of velocities under this transformation \cite{seifert2012stochastic}. Using the law of total probability combined with the Onsager-Machlup \cite{onsager1953fluctuations} form of the positional path probability in the Stratonovich discretisation, we find
\begin{align}
    \mathbb{P}[\{x\}_0^T] &= \int \mathcal{D}w \ \mathbb{P}[\{x\}_0^T | \{w\}_0^T] \mathbb{P}[\{w\}_0^T] \nonumber \\
    &\propto \int \mathcal{D}w \ {\rm exp}\left( - \frac{1}{4D_x} \int_0^T dt \ (\dot{x}(t) - w(t))^{2} \right)\mathbb{P}[\{w\}_0^T]~, \label{eq:OM_tfor}
\end{align}
and similarly
\begin{align}
    \mathbb{P}[\{x^R\}_0^T] 
    &\propto \int \mathcal{D}w \ {\rm exp}\left( - \frac{1}{4D_x} \int_0^T dt \ (\dot{x}(t) + w^R(t))^{2} \right)\mathbb{P}[\{w\}_0^T]~. \label{eq:OM_trev}
\end{align} 
Assuming PT symmetry of the $w$-statistics is equivalent to requiring that $\mathbb{P}[\{w\}_0^T] = \mathbb{P}[\{-w^R\}_0^T]$. Substituting into Eq.~\eqref{eq:OM_trev} gives 
\begin{align}
    \mathbb{P}[\{x^R\}_0^T] 
    &\propto \int \mathcal{D}w \ {\rm exp}\left( - \frac{1}{4D_x} \int_0^T dt \ (\dot{x}(t) + w^R(t))^{ 2} \right)\mathbb{P}[\{-w^R\}_0^T] \nonumber \\
    &\propto \int \mathcal{D}\tilde{w} \ {\rm exp}\left( - \frac{1}{4D_x} \int_0^T dt \ (\dot{x}(t) - \tilde{w}(t))^{ 2} \right)\mathbb{P}[\{\tilde{w}\}_0^T]~, \label{eq:trev_reex}
\end{align}
where we have used 
\begin{equation}
    \int \mathcal{D}w \ F[w^R(t)]= \int \mathcal{D}(-w^R) \ F[w^R(t)] = \int \mathcal{D}\tilde{w} \ F[- \tilde{w}(t)]~.
\end{equation}
Combining Eqs.~\eqref{eq:OM_tfor} and \eqref{eq:trev_reex} with Eq.~\eqref{eq:si_inf}, we find $\Delta S[\{x\}_0^T] = 0$, confirming the positional dynamics of the active particle are indeed time-reversal symmetric.
It is worth restating that PT symmetry is a weaker constraint on the $w$-statistics compared to requiring that symmetry under parity reversal ($\mathbb{P}[\{w\}_0^T] = \mathbb{P}[\{-w\}_0^T]$) and time reversal ($\mathbb{P}[\{w\}_0^T] = \mathbb{P}[\{w^R\}_0^T]$) are separately satisfied. The simplest example of a stochastic process that is PT symmetric, but not P \emph{or} T symmetric, is the zero-mean three-state Markov process taking values $w \in \{-1,0,1\}$ with transition rates $k_{0,1} = k_{1,-1} = k_{-1,0} = \alpha$ and $k_{1,0} = k_{-1,1} = k_{0,-1} = \beta$, where $k_{i,j}$ denotes the probability per unit time to transition from state $j$ into state $i$ and $\alpha \neq \beta$ are positive real numbers.

\section{Conditional correlations of the symmetric telegraph process} \label{sr:II}
In this supplementary section, we derive the n-time correlation functions for a normalised telegraph process with symmetric transition rate $\alpha$ between states $w \in \{-1,+1\}$, conditioned on the process achieving a given state at some final time $T$ (cf.\ the unconditional correlations derived in Appendix A of Ref.~\cite{medeiros2021effective}). In particular, given an ordered set of measurement times $0 < t_1 < t_2 < ... < t_n \leq T$, we denote
\begin{subequations}
\begin{alignat}{2}
    \overline{w(t_1) w(t_2)...w(t_n)}^{(+1)} &= \int \mathcal{D}w \ w(t_1)...w(t_n) \mathbb{P}(\{w\}_0^T|w(T)=+1) \nonumber \\
    &= \int \mathcal{D}w' \ w'(T-t_n)...w'(T-t_1) \mathbb{P}(\{w'\}_0^T|w'(0)=+1) \label{eq:rarr_1}\\
    \overline{w(t_1) w(t_2)...w(t_n)}^{(-1)} &= \int \mathcal{D}w \ w(t_1)...w(t_n)  \mathbb{P}(\{w\}_0^T|w(T)=-1) \nonumber \\
    &= \int \mathcal{D}w' \ w'(T-t_n)...w'(T-t_1) \mathbb{P}(\{w'\}_0^T|w'(0)=-1) \label{eq:rarr_2}
\end{alignat}
\end{subequations}
where, in the second equality of both Eqs.~\eqref{eq:rarr_1} and \eqref{eq:rarr_2}, we have used the invariance of the symmetric telegraph process statistics under time reversal to express the expectations in terms of conditioning on some initial state $w(0)$ of a trajectory of duration $T$. Let us now introduce the transition probabilities
\begin{subequations}
\begin{alignat}{2}
    P_{+1,+1}(t) = \frac{1}{2} \left( 1 + e^{-2\alpha t}\right) \\
    P_{-1,+1}(t) = \frac{1}{2} \left( 1 - e^{-2\alpha t}\right) \\
    P_{+1,-1}(t) = \frac{1}{2} \left( 1 - e^{-2\alpha t}\right) \\
    P_{-1,-1}(t) = \frac{1}{2} \left( 1 + e^{-2\alpha t}\right)
\end{alignat}
\end{subequations}
where $P_{i,j}(t)$ denotes the probability of a telegraph process initialised in state $j$ to be found in state $i$ after time $t$. Defining the matrix
\begin{equation}
    M_w(t) = 
    \begin{bmatrix}
    (+1) P_{+1,+1}(t) & (+1) P_{+1,-1}(t) \\
    (-1) P_{-1,+1}(t) & (-1) P_{-1,-1}(t)
\end{bmatrix}~,
\end{equation}
we can conveniently reexpress the time-reversed conditional correlation functions above as 
\begin{subequations}
\begin{alignat}{2}
    \overline{w(t_1) w(t_2)...w(t_n)}^{(+1)} &= 
    \begin{bmatrix}
    1 \\
    1
    \end{bmatrix}^T
M_w(t_2-t_1) M_w(t_3-t_2)...M_w(T-t_n)
 \begin{bmatrix}
    1 \\
    0
    \end{bmatrix} \\
    \overline{w(t_1) w(t_2)...w(t_n)}^{(-1)} &= 
    \begin{bmatrix}
    1 \\
    1
    \end{bmatrix}^T
M_w(t_2-t_1) M_w(t_3-t_2)...M_w(T-t_n)
 \begin{bmatrix}
    0 \\
    1
    \end{bmatrix}~,
\end{alignat}
\end{subequations}
which only differ by the final column vector, responsible for the conditioning.
For the particular case of a symmetric telegraph process,
\begin{equation}
    \begin{bmatrix}
    1 \\
    1
    \end{bmatrix}^T M_w(t_2 - t_1) = e^{-2\alpha(t_2-t_1)} 
    \begin{bmatrix}
    1 \\
    -1
    \end{bmatrix}^T~,
\end{equation}
and therefore
\begin{equation}
    \begin{bmatrix}
    1 \\
    1
    \end{bmatrix}^T M_w(t_2 - t_1) M_w(t_3-t_2) = e^{-2\alpha(t_2-t_1)} 
    \begin{bmatrix}
    1 \\
    1
    \end{bmatrix}^T~,
\end{equation}
whence we finally obtain
\begin{equation}
    \overline{w(t_1) w(t_2)...w(t_n)}^{(+1)} = 
    \begin{cases}
    \prod_{m=1}^{n_{\%2}} e^{-2\alpha(t_{t_{2m}-t_{2m-1}})}, & \text{for } n {\rm \ even} \\
    e^{-2\alpha(T-t_n)}\prod_{m}^{n//2} e^{-2\alpha(t_{t_{2m}-t_{2m-1}})}, & \text{for } n {\rm \ odd} 
  \end{cases}~,
\end{equation}
while 
\begin{equation}
    \overline{w(t_1) w(t_2)...w(t_n)}^{(-1)} = 
    \begin{cases}
    \prod_{m=1}^{n_{\%2}} e^{-2\alpha(t_{t_{2m}-t_{2m-1}})}, & \text{for } n {\rm \ even} \\
    -e^{-2\alpha(T-t_n)}\prod_{m}^{n//2} e^{-2\alpha(t_{t_{2m}-t_{2m-1}})}, & \text{for } n {\rm \ odd} 
  \end{cases}~,
\end{equation}
where $n_{\%2}$ denotes integer division by $2$.
The lowest-order correlation functions are thus given by
\begin{align}
    \overline{w(t_1)}^{(+1)} = e^{-2\alpha(T-t_1)}~, \quad \overline{w(t_1)w(t_2)}^{(+1)} = e^{-2\alpha(t_2-t_1)}~, \quad \overline{w(t_1)w(t_2)w(t_3)}^{(+1)} = e^{-2\alpha(t_2-t_1)} e^{-2\alpha(T-t_3)}~.
\end{align}
The odd cumulants up to the same order are, instead,
\begin{subequations}
\begin{alignat}{2}
    \overline{w(t_1)}^{(+1),c} &= e^{-2\alpha(T-t_1)} \\
    \overline{w(t_1) w(t_2) w(t_3)}^{(+1),c} &= -4 e^{-2\alpha(T-t_1)}e^{-2\alpha(T-t_2)}\sinh[2\alpha(T-t_3)]~.
\end{alignat}
\end{subequations}
Note that, by symmetry of the $w$-statistics under parity reversal, $w(t) \to - w(t)$, we have 
\begin{equation}
    \overline{w(t_1) w(t_2) ... w(t_n)}^{(+1),c} - \overline{w(t_1) w(t_2) ... w(t_n)}^{(-1),c} = 
    \begin{cases}
    0,  &\text{if $n$ even} \\
    2 \overline{w(t_1) ... w(t_n)}^{(+1),c}  &\text{if $n$ odd}
    \end{cases}~. \label{eq:symm_conn_mom}
    \end{equation}

\section{Moments of position for active particles in harmonic potentials} \label{sr:III}
In this supplementary section we derive simple recurrence formulas for the moments of the position of RnT and AOU particles in harmonic potentials. The relevance of these results for the present work originates from the mapping we identified in the main text, via Eq.~(\ref{eq:Q_leading}), between the dynamics of the fluctuating extracted power from a RnT particle upon application of the optimal protocol $F_{\rm ext}^*(t) = -{\mathbb{E}_w[ w (T) | \{x\}_0^T ]}/{2}$ and those of the position of such a harmonically-bound active particle. We will see that a similar mapping applies to the AOU particle, see Eq.~\eqref{eq:G_conv_aoup} below.  While the exact form of the full probability densities are typically quite cumbersome (see, for example, Ref.~\cite{garcia2021run}), these recurrence formulas allow us to easily extract key characteristics such as the (positive) mean power output and the associated coefficient of variation, defined as the standard deviation normalised to the mean.

\subsection{RnT particle}
We now derive the recurrence formulas for the steady-state positional moments of an RnT particle, which draw on the steady-state positional moments conditioned on the particle being in either of its accessible self-propelling modes. We arbitrarily choose to calculate the positional moments of the right-moving state ($w(t) = +\nu$), since those of the left-moving state ($w(t) = -\nu$) can be obtained by symmetry, assuming as we do here that the switching rates are symmetric. We start from the coupled Fokker-Planck equations for the joint probability density $\mathcal{P}_{+/-}(x,t)$ to observe a particle in the right/left-moving state at position $x$ at time $t$,
\begin{subequations}\label{eq:fp_RnT_harm}
\begin{alignat}{2}
    \partial_t \mathcal{P}_{+} &= D_x \partial_x^2 \mathcal{P}_{+} + \kappa \partial_x (x \mathcal{P}_{+}) - \nu \partial_x \mathcal{P}_{+} + \alpha\left(\mathcal{P}_{-} - \mathcal{P}_{+}\right) \label{eq:fp_RnT_harm_RightMover}~,\\
    \partial_t \mathcal{P}_{-} &= D_x \partial_x^2 \mathcal{P}_{-} + \kappa \partial_x (x \mathcal{P}_{-}) + \nu \partial_x \mathcal{P}_{-} - \alpha\left(\mathcal{P}_{-} - \mathcal{P}_{+}\right) ~,
\end{alignat}
\end{subequations}
where $\kappa$ denotes the stiffness of the harmonic potential and we retain all other definitions of symbols from the main text.
Since, at steady-state, $\mathcal{P}_{+}(x,t) = \mathcal{P}_{-}(-x,t)$ by symmetry, the final term in Eq.~(\ref{eq:fp_RnT_harm_RightMover}) is odd. 
It can be written equivalently as $\alpha\left(\mathcal{P}_{-} - \mathcal{P}_{+}\right) = \alpha\left(\mathcal{P} - 2\mathcal{P}_{+}\right)$ where the total density $\mathcal{P} = \mathcal{P}_{+} + \mathcal{P}_{-}$ is an even function of $x$.
At steady-state $\partial_{t}\mathcal{P}_{+}=\partial_{t}\mathcal{P}_{-}=0$, and after operating on Eq.~(\ref{eq:fp_RnT_harm_RightMover}) with $\int dx~x^{n}$, performing all integrations by parts, and exploiting the even/odd symmetry of the final term, we eventually arrive at
\begin{equation} \label{eq:FP_moments}
0 =
\begin{cases}
D_{x} n (n-1)\mathbb{E}_{x}^{+}[x^{n-2}] - \kappa n \mathbb{E}_{x}^{+}[x^{n}] + \nu n \mathbb{E}_{x}^{+}[x^{n-1}] & \text{for even } n \geq 2 \\
D_{x} n (n-1)\mathbb{E}_{x}^{+}[x^{n-2}] - \kappa n \mathbb{E}_{x}^{+}[x^{n}] + \nu n \mathbb{E}_{x}^{+}[x^{n-1}] - 2\alpha \mathbb{E}_{x}^{+}[x^{n}] & \text{for odd } n \geq 1
\end{cases}~,
\end{equation}
where we have defined the steady-state conditional moments $\mathbb{E}_{x}^{+}[x^{n}] \equiv \lim_{t \to \infty}\int dx~x^{n}\mathcal{P}_{+}(x,t)$ and used that all boundary terms vanish due to $\lim_{x \to \pm\infty} \mathcal{P}_{+}(x,t)= 0$. Eq.~(\ref{eq:FP_moments}) can be rearranged into the form of a recursive relation for the $n^{\rm th}$ conditional moment,
\begin{equation} \label{eq:moments_RnT_harmonic}
\mathbb{E}_{x}^{+}[x^{n}] =
\begin{cases}
\frac{1}{\kappa}\left(D_{x}(n-1)\mathbb{E}_{x}^{+}[x^{n-2}] + \nu\mathbb{E}_{x}^{+}[x^{n-1}]\right) & \text{for even } n \geq 2 \\
\frac{n}{n \kappa + 2\alpha} \left( D_{x}(n-1)\mathbb{E}_{x}^{+}[x^{n-2}] + \nu \mathbb{E}_{x}^{+}[x^{n-1}] \right) & \text{for odd } n \geq 1
\end{cases}~.
\end{equation}
Using Eq.~\eqref{eq:moments_RnT_harmonic}, the normalisation condition $\mathbb{E}_{x}^{+}[x^{0}] = \int dx~\mathcal{P}_{+}(x) = 1/2$ readily gives rise to the $n=1$ moment, $\mathbb{E}_{x}^{+}[x] = \nu/(2(\kappa + 2\alpha))$ which, in turn, can be used to obtain the $n=2$ moment, $\mathbb{E}_{x}^{+}[x^{2}] = D_{x}/2\kappa + \nu^{2}/(2\kappa(\kappa+2\alpha))$, in agreement with Ref.~\cite{garcia2021run}, with care being taken to adjust for the different definition of the `tumble' rate $\alpha$ used in that work. In fact, closed-form expressions for all higher-order conditional moments can be obtained in a similar manner, since any given conditional moment depends only on the two lower-order moments immediately preceding it.

The steady-state conditional moments for the left-moving state, denoted $\mathbb{E}_{x}^{-}[x^{n}] \equiv \lim_{t \to \infty}\int dx~x^{n}\mathcal{P}_{-}(x,t)$, and those of the total density $\mathcal{P}$ are readily obtained from the symmetry relation $\mathbb{E}_{x}^{-}[x^{n}] = (-1)^{n} \mathbb{E}_{x}^{+}[x^{n}]$ and the total probability relation $\mathbb{E}_{x}[x^{n}] = \mathbb{E}_{x}^{+}[x^{n}] + \mathbb{E}_{x}^{-}[x^{n}]$, respectively. It is thus straightforward to obtain the expression for the variance of the particle position,
\begin{equation}\label{eq:RnT_mom2}
    \mathbb{E}_x[x^2] = \frac{D_{x}}{\kappa} + \frac{\nu^{2}}{\kappa(\kappa+2\alpha)}~.
\end{equation}
As was shown in the main text, in the asymptotic regime ${\rm Pe} \ll 1$, the dynamics of the fluctuating power extracted from an RnT process with hidden self-propulsion state $w(t)$ subject to the optimal protocol $F_{\rm ext}^*$ are identical to those of the square of the position of an RnT particle in a harmonic potential. There, we then used our result for the second moment of the latter, Eq.~\eqref{eq:RnT_mom2}, to compute the average power output, which defines the upper bound to the average extractable work in this regime. Remarkably, the same correspondence can be exploited to relate the fourth moment of the particle position,
\begin{align}
    \mathbb{E}_x[x^4]
    = \frac{3D_{x}^{2}}{\kappa^{2}} + \frac{6D_{x} \nu^{2}}{\kappa^{2}(\kappa + 2\alpha)} + \frac{3\nu^{4}}{\kappa^{2}(\kappa + 2\alpha)(3\kappa + 2\alpha)}~,
    \label{eq:RnT_mom4}
\end{align}
to the second moment of the power output distribution upon application of the optimal protocol $F_{\rm ext}^*$, which is found to be
\begin{equation}
    \mathbb{E}_{\xi,w}[\dot{W}^2_{\rm RnT}(F^*_{\rm ext})] \simeq 
    \left( \frac{\nu^2 }{4} \right)^2 \frac{\mathbb{E}_{\xi,w}[ (Q[\{x\}_0^T])^4]}{16}
    = \left( \frac{\nu^4 }{16} \right) \cdot \frac{3}{64} {\rm Pe}^2 + \mathcal{O}({\rm Pe}^3)
\end{equation}
to leading order in small Pe. Correspondingly, the coefficient of variation in this regime is 
\begin{equation} \label{eq:cov_rnt}
    C^{W}_{\rm RnT} \equiv 
    \frac{\sqrt{\mathbb{E}_{\xi,w}[\dot{W}^2_{\rm RnT}(F^*_{\rm ext})] - (\mathbb{E}_{\xi,w}[\dot{W}_{\rm RnT}(F^*_{\rm ext})])^2}}{\mathbb{E}_{\xi,w}[\dot{W}_{\rm RnT}(F^*_{\rm ext})]} = \sqrt{2} = 1.414...
\end{equation}
indicating that, even under optimal control, fluctuations often lead to transients of negative power extaction. Eq.~\eqref{eq:RnT_mom4} correctly reduces to the Gaussian kurtosis $\mathbb{E}_x[x^4]/(\mathbb{E}_x[x^2])^2 = 3$ in the limit $\alpha \to \infty$, where the self-propulsion becomes dynamically (although not thermodynamically \cite{celani2012anomalous,cocconi2020entropy}) irrelevant.

\subsection{AOU particle}
We now discuss the derivation of the moments of the positional probability density for an active Ornstein-Uhlenbeck process \cite{martin2021statistical} in a harmonic potential.
First, we calculate the second moment of the particle position $x(t)$, which is used in Sec.~\ref{sr:V} below to derive an upper bound to the average extractable work in the asymptotic regime ${\rm Pe} \ll 1$, Eq.~(\ref{eq:asym_lowPe-Aou_SM}). We start from the Fokker-Planck equation for the joint probability density $\mathcal{P}(x,w,t)$ that the particle is at position $x$ exhibiting an instantaneous self-propulsion velocity $w(t)$ at time $t$,
\begin{equation} \label{eq:fp_aoup_harm}
    \partial_t \mathcal{P} = D_x \partial_x^2 \mathcal{P} + \kappa \partial_x (x \mathcal{P}) - w \partial_x \mathcal{P} + D_w \partial_w^2 \mathcal{P} + \mu \partial_w (w \mathcal{P}) ~,
\end{equation}
where $\kappa$ denotes the stiffness of the harmonic potential acting on the position $x$ and $\mu$ 
that of the harmonic potential acting on the self-propulsion velocity $w(t)$. At steady state, $\partial_t \mathcal{P}(x,w,t)=0$, we multiply the right-hand side of Eq.~\eqref{eq:fp_aoup_harm} by $x^2$ and $xw$ and integrate with respect to both $x$ and $w$ to obtain, respectively, 
\begin{subequations} \label{eq:mom11_aou}
\begin{alignat}{2}
    0 &= D_x - \kappa \mathbb{E}_{x,w}[ x^2 ] + \mathbb{E}_{x,w}[ x w ]~, \\
    0 &= -(\kappa + \mu) \mathbb{E}_{x,w}[ x w ] + \mathbb{E}_{x,w}[ w^2 ] ~.
\end{alignat}
\end{subequations}
Combining the equations above with the known second moment of the standard Ornstein-Uhlenbeck process \cite{gardiner1985handbook}, $\mathbb{E}_{x,w}[ w^2 ] = D_w/\mu$, we eventually arrive at the desired result
\begin{equation} \label{eq:AOUP_mom2}
    \mathbb{E}_{x,w}[ x^2 ] = \frac{D_x}{\kappa} + \frac{D_w}{\mu \kappa (\mu + \kappa)}~.
\end{equation}
Eq.~\eqref{eq:AOUP_mom2} is consistent with the variance of $x$ reducing to that of a standard OU process in the equilibrium limit $D_w = 0$, where $\mathbb{E}_{x,w}[ x^{2} ] = D_x/\kappa$.
A similar procedure can be followed to calculate higher-order moments of $x$ given that the moments $\mathbb{E}_{x,w}[w^k]$ are known from the literature on the OU process \cite{gardiner1985handbook},
\begin{equation}\label{eq:AOUP_VelocityMoments}
    \mathbb{E}_{x,w}[w^k] = 
    \begin{cases}
    0& \text{if $k$ is odd}  \\
    (k-1)!! (D_w/\mu)^\frac{k}{2} & \text{if $k$ is even}
    \end{cases} ~.
\end{equation}
In particular, we start from the Fokker-Planck equation \eqref{eq:fp_aoup_harm} at steady state and multiply by $x^n w^m$ ($n,m \in \mathbb{N}$) before integrating with respect to both variables to obtain
\begin{equation}\label{eq:sys_xw_mom_aou}
\begin{small}
0 = D_x (n)_{2} \mathbb{E}_{x,w}[x^{n-2} w^m] - (\kappa n + \mu m) \mathbb{E}_{x,w}[x^n w^m] + n \mathbb{E}_{x,w}[x^{n-1} w^{m+1}] + D_w (m)_{2} \mathbb{E}_{x,w}[x^n w^{m-2}]~,
\end{small}
\end{equation}
for $ n,m \geq 0$, defining the falling factorial $(k)_{2} = k(k-1)$ for compactness. Together with Eq.~\eqref{eq:mom11_aou}, the system of equations \eqref{eq:sys_xw_mom_aou} can be solved to compute any desired moment $\mathbb{E}_{x,w}[x^n]$. Here, only even moments are non-trivial since $\mathbb{E}_{x,w}[x^{2k+1}]=0$ for all $k \in \mathbb{N}$ by symmetry. Similarly, we expect $\mathbb{E}_{x,w}[x^{2k+1} w^{2\ell}]=\mathbb{E}_{x,w}[x^{2k} w^{2\ell+1}]=0$ with $\ell \in \mathbb{N}$. Combining the expressions of the type \eqref{eq:sys_xw_mom_aou} obtained from all choices of $n$ and $m$ such that $n + m = 4$ with Eqs.~(\ref{eq:mom11_aou})-(\ref{eq:AOUP_VelocityMoments}), we eventually find the Gaussian kurtosis
\begin{align}\label{eq:AOUPmom4}
    \mathbb{E}_{x,w}[x^4] = 3 \left(\mathbb{E}_{x,w}[x^2]\right)^2~.
\end{align}

As shown in SM Sec.~\ref{sr:V} below, in the asymptotic regime ${\rm Pe} \ll 1$, the dynamics of the fluctuating power extracted from an AOU process with hidden self-propulsion state $w(t)$ subject to the optimal protocol $F_{\rm ext}^*$ are identical to those of the square of the position of an AOU particle in a harmonic potential. There, we use our result for the second moment of the latter, Eq.~\eqref{eq:AOUP_mom2}, to compute the average power output, which defines the upper bound to the average extractable work in this regime, Eq.~\eqref{eq:asym_lowPe-Aou_SM}. The same correspondence allows us to compute the second moment of the power output through Eq.~\eqref{eq:AOUPmom4} to leading order in small Pe,
\begin{equation}
    \mathbb{E}_{\xi,w}[\dot{W}^2_{\rm AOU}(F^*_{\rm ext})] \simeq 
    \left( \frac{\sigma_w^2 }{4} \right)^2 \mathbb{E}_{\xi,w}[ (\mathcal{L}^{(1)}[\{x\}_0^T])^4]
    = \left( \frac{\sigma_w^4 }{16} \right) \cdot \frac{3}{2^{10}}\rm Pe^2 + \mathcal{O}({\rm Pe}^3)~.
\end{equation}
The coefficient of variation in this regime is thus
\begin{equation}
    C^{W}_{\rm AOU} \equiv \frac{\sqrt{\mathbb{E}_{\xi,w}[\dot{W}^2_{\rm AOU}(F^*_{\rm ext})] - (\mathbb{E}_{\xi,w}[\dot{W}_{\rm AOU}(F^*_{\rm ext})])^2}}{\mathbb{E}_{\xi,w}[\dot{W}_{\rm AOU}(F^*_{\rm ext})]} = \sqrt{2} = 1.414...
\end{equation}
i.e.\ the same as for the RnT case in the previous section, Eq.~\eqref{eq:cov_rnt}.

\section{Optimal protocol for an AOU particle at ${\rm Pe}\ll 1$}\label{sr:V}
In this supplementary section, we illustrate the general result for the optimal protocol at low Pe obtained in the main text, Eq.~(\ref{eq:mathcalL_exp_one}), by applying it to the specific case of a one-dimensional AOU process, the simplest canonical active particle model with a continuous self-propulsion state \cite{martin2021statistical}. The dynamics of the self-propulsion velocity $w(t)$ are captured by the equilibrium linear Langevin equation $\dot{w}(t) = - \mu w(t) + \sqrt{2 D_w} \eta(t)$, with $\eta(t)$ a zero-mean, unit variance white noise with diffusivity $D_w$. The prior probability density is thus Boltzmann, $\mathbb{P}(w) \propto {\rm exp}(-\mu w^2/(2D_w))$, whence $\sigma_w^2 = D_w/\mu$. Due to $w(t)$ being a Gaussian process we have that $ \mathcal{L}[\{x\}_0^T,v] \equiv \mathcal{L}^{(0)}[\{x\}_0^T] + (v/\sigma_w) \mathcal{L}^{(1)}[\{x\}_0^T]$, i.e.\ conditional cumulants are, at most, linear in $v$ \cite{gardiner1985handbook}. Eq.~(\ref{eq:v_opt_gen_L}) thus reduces to
\begin{equation} \label{eq:aou_expv_momgen_SM}
    \mathbb{E}_w[w(T) | \{x\}_0^T] = \left. \frac{\partial \mathcal{H}_{\rm AOU}(z)}{\partial z}  \right|_{z = \mathcal{L}^{(1)}/\sigma_w}~,
\end{equation}
where $\mathcal{H}_{\rm AOU}(z) = \log \mathbb{E}_w[ e^{w z }] = \sigma_w^2 z^2/2$ is the known cumulant-generating function of the fluctuating self-propulsion velocity prior probability density $\mathbb{P}(w)$. 
Combining Eqs.~(\ref{eq:opt_force_general}) and \eqref{eq:aou_expv_momgen_SM}, we find the optimal protocol
\begin{equation}
    F_{\rm ext}^*(t) = - \frac{\sigma_w}{2} \mathcal{L}^{(1)}[\{x\}_0^T]
\end{equation}
and thus
\begin{equation}
    \mathbb{E}_{\xi,w}[\dot{W}_{\rm AOU}(F^*_{\rm ext})] = \frac{\sigma_w^2}{4}   \mathbb{E}_{\xi,w}\big[\mathcal{L}^{(1)}[\{x\}_0^T]^2\big]~.
\end{equation}
At this point, the only outstanding challenge is to compute the dimensionless functional $\mathcal{L}^{(1)}[\{x\}_0^T]$. To leading order in ${\rm Pe} \ll 1$,
\begin{equation}
    \mathcal{L}^{(1)}[\{x\}_0^T] = \frac{\rm Pe}{4} \int_0^T dt \left( \frac{\mu \dot{x}_c(t)}{\sigma_w } \right) e^{- \mu (T-t)} + \mathcal{O}({\rm Pe}^2)~, \label{eq:G_conv_aoup}
\end{equation}
where we have used $\overline{w(t)}^{(v),c} = v \exp(- \mu (T-t))$. 
The same result can be obtained starting from Eq.~(\ref{eq:mathcalL_exp_one}), noticing the second conditional cumulant of the OU process, $\overline{w^2(t)}^{(v),c}$, is independent of $v$. By rescaling $T'=\mu T$ as in the RnT case, Eq.~(\ref{eq:Qeq_resc}), and taking $\dot{x}_c(t) = 0$ for $t<0$, we can recast Eq.~\eqref{eq:G_conv_aoup} into a differential equation for $\mathcal{L}^{(1)}(T')$ mirroring that of an AOU particle in a harmonic potential,
\begin{equation}
    \frac{d{\mathcal{L}}^{(1)}(T')}{dT'} = \frac{{\rm Pe}}{4} \cdot \frac{w(T')}{\sigma_w} - \mathcal{L}^{(1)}(T') + \sqrt{\frac{{\rm Pe}}{8}} \cdot \eta(T')~.
\end{equation}
This mapping allows us to extract a tight upper bound to the average extractable power from an AOU particle with hidden self-propulsion velocity in the low-${\rm Pe}$ asymptote,
\begin{equation} \label{eq:asym_lowPe-Aou_SM}
    \mathbb{E}_{\xi,w}[\dot{W}_{\rm AOU}(F^*_{\rm ext})] = \frac{\sigma_w^2 }{4} \frac{\rm Pe}{16} + \mathcal{O}({\rm Pe}^2)~.
\end{equation}

\section{Conditional expectations from empirical measurements} \label{sr:new_infer}
Consider the overdamped Langevin equation for a generic active particle, $\dot{x}(t)=w(t) +\xi(t)$ with $\langle \xi(t) \xi(t')\rangle = 2D_x \delta(t-t')$, where $w(t)$ represents a hidden self-propulsion velocity governed by an unknown Langevin or master equation. In this supplementary section, we outline a procedure to relate expectations of any function $\bullet = \bullet(\dot{x})$ of the net velocity \emph{conditioned on a particular final net velocity} $\dot{x}(T)$, to correlation functions of the same observable \emph{conditioned on a particular final value $w(T)$ of the hidden self-propulsion velocity}. While the first type of expectation is empirically accessible, it is the second which appears in the definition of the generic optimal protocol at low $\rm Pe$, Eq.~(\ref{eq:mathcalL_exp_one}) in the main text. The ability to derive the latter from the former is thus key to achieving maximum power output even when the Langevin/master equation governing the self-propulsion velocity is not known \textit{a priori}.
Importantly, we assume the statistics of the additive noise (here the value of the diffusivity $D_x$) are known, e.g.\ by characterising the dynamics of the particle in an experimental condition where activity is inhibited or by considering a passive probe with the physical properties. 

We first introduce the notation for the two types of conditional expectations, 
\begin{align}
    \overline{\bullet}^{(v)} &\equiv \int \mathcal{D}w \ \bullet \mathbb{P}_w[ \{w(t)\}_0^T| w(T)=v] \label{eq:cond_corr_w}\\
    \widebarbar{\bullet}^{(v)} &\equiv \int \mathcal{D}x \ \bullet \mathbb{P}_{\dot{x}}
[ \{\dot{x}(t)\}_0^T| \dot{x}(T)=v]~. \label{eq:cond_corr_x}
\end{align}
Here, we use $\delta(\cdot)$ and $\delta[\cdot]$ to denote Dirac delta functions and functionals, respectively. Similarly, we distinguish between the steady-state probability density function and path probability density for a given stochastic process $y$ by denoting them as $P_y(\cdot)$ and $\mathbb{P}_y[\cdot]$ respectively. With this notation in place, conditional correlations of the type \eqref{eq:cond_corr_x} can be related to those of the type \eqref{eq:cond_corr_w} as follows,
\begin{subequations}
\begin{align}
    \widebarbar{\bullet}^{(v)}
    &= \frac{1}{P_{\dot{x}}(v)}  \int \mathcal{D}x \bullet \mathbb{P}_{\dot{x}}[\{\dot{x}\}_0^T] \delta(\dot{x}(T)-v)  \label{eq:st1}\\
    &= \frac{1}{P_{\dot{x}}(v)} \int \mathcal{D}x \bullet \delta(\dot{x}(T)-v) \int \mathcal{D}\xi \ \mathbb{P}_\xi[\{\xi\}_0^T] \mathbb{P}_w[\{\dot{x}-\xi\}_0^T] \label{eq:st2} \\
    &= \frac{1}{P_{\dot{x}}(v)} \int \mathcal{D}x \bullet \delta(\dot{x}(T)-v) \int \mathcal{D}\xi \mathcal{D}w \ \mathbb{P}_\xi[\{\xi\}_0^T] \mathbb{P}_w[\{w\}_0^T] \delta[w+\xi-\dot{x}]  \label{eq:st3}\\
    &= \frac{1}{P_{\dot{x}}(v)} \int \mathcal{D}w \mathcal{D}\xi \ \mathbb{P}_\xi[\{\xi\}_0^T] \mathbb{P}_w[\{w\}_0^T] \int \mathcal{D}x \ \bullet \delta(\dot{x}(T)-v) \delta[\omega +\xi-\dot{x}] \label{eq:st4} \\
    &= \frac{1}{P_{\dot{x}}(v)} \int \mathcal{D}\xi \ \mathbb{P}_\xi[\{\xi\}_0^T] \int \mathcal{D}w \ \mathbb{P}_w[\{w\}_0^T] \bullet|_{\dot{x}=w+\xi} \delta(w(T)+\xi(T)-v) \label{eq:st5}\\
    &= \frac{1}{P_{\dot{x}}(v)} \int \mathcal{D}\xi \ \mathbb{P}_\xi[\{\xi\}_0^T] \overline{\bullet}^{(v-\xi(T))} P_w(v-\xi(T))~, \label{eq:st6}
\end{align}
\end{subequations}
where in going from \eqref{eq:st1} to \eqref{eq:st2} we have used that the path probability of $\dot{x}(t)=w(t)+\xi(t)$ is the convolution of the path probabilities of the summands,
\begin{equation}
    \mathbb{P}_{\dot{x}}[\{\dot{x}=w+\xi\}_0^T] = \int \mathcal{D}\xi \ \mathbb{P}_\xi[\{\xi\}_0^T] \mathbb{P}_w[\{\dot{x}-\xi\}_0^T]~,
\end{equation}
while in going from \eqref{eq:st2} to \eqref{eq:st3} we have used the property of the Dirac delta functional,
\begin{equation}
    \mathbb{P}_w[\{\dot{x}-\xi\}_0^T] = \int \mathcal{D}w \ \mathbb{P}_w[\{w\}_0^T] \delta[w+\xi-\dot{x}]~.
\end{equation}
Finally, in going from \eqref{eq:st5} to \eqref{eq:st6} we have used \eqref{eq:cond_corr_w} to write the functional integral over the self-propulsion velocity path $w(t)$ in terms of the expectation conditioned on a particular final value. The probability densities $P_{\dot{x}}$ and $P_{w}$ appearing in \eqref{eq:st6} can both be extracted from observable trajectories: the first by simply constructing empirical histograms of the net velocity $\dot{x}$ and the second by deconvolution of the relation $P_{\dot{x}}(v) = \int d\xi P_\xi(\xi) P_w(\dot{x}-\xi)$ originating from the additive nature of the noise, where again we assume $ P_\xi$ to be known. Note the definition of $P_\xi$ for the case of white noise requires a suitable time discretisation of the recorder trajectories, as discussed below. \\

Once again, our aim is to express the conditional correlations $\overline{w(t)}^{(v)}$ and $\overline{w^2(t)}^{(v)}$ appearing in the definition (22) of the generic optimal protocol at low Pecl{\'e}t number in terms of correlations that can be measured empirically. Equipped with Eq.~\eqref{eq:st6}, we can then consider particular choices for the observable $\bullet(\dot{x})$. Firstly, let $\bullet = \dot{x}(t)$ with $t < T$,
\begin{align}
    \widebarbar{\dot{x}(t)}^{(v)} P_{\dot{x}}(v) &= \int \mathcal{D}\xi \ \mathbb{P}_\xi[\{\xi\}_0^T] \left[ \overline{w(t) + \xi(t)}^{(v-\xi(T))} \right] P_w(v-\xi(T)) \nonumber \\
    &= \int \mathcal{D}\xi \ \mathbb{P}_\xi[\{\xi\}_0^T]  \overline{w(t)}^{(v-\xi(T))} P_w(v-\xi(T))~,\label{eq:conv_o1}
\end{align}
where we have used linearity of the conditional expectation together with the fact that $\xi$ is zero-mean. Secondly, let $\bullet = \dot{x}^2(t)$ with $t<T$,
\begin{align}
    \widebarbar{\dot{x}^2(t)}^{(v)} P_{\dot{x}}(v) &= \int \mathcal{D}\xi \ \mathbb{P}_\xi[\{\xi\}_0^T] \left[ \overline{w^2(t) + \xi^2(t) + 2\xi(t)w(t)}^{(v-\xi(T))} \right] P_w(v-\xi(T)) \nonumber \\
    &= \int \mathcal{D}\xi \ \mathbb{P}_\xi[\{\xi\}_0^T]  \overline{w^2(t)}^{(v-\xi(T))}  P_w(v-\xi(T)) + \int \mathcal{D}\xi \ \mathbb{P}_\xi[\{\xi\}_0^T]   \xi^2(t) P_w(v-\xi(T))~, \label{eq:conv_o2}
\end{align}
where again we have used that $\xi$ is zero mean to remove the cross term. For the last step and in order to regularise the high frequency behaviour of the white noise $\xi$, we are required to introduce a time discretisation of the empirical trajectories with timestep $\tau = T/N$, which we take to be small compared to all other characteristic timescales of the problem. We thus express functional integrals over noise realisations as products of regular integrals
\begin{equation}\label{eq:discr_path_int}
    \int \mathcal{D}\xi \ \mathbb{P}_\xi[\{\xi\}_0^T] = \int \prod_{i=0}^N d\xi_i P_\xi(\xi_i)~,
\end{equation}
where the subscript $i=0,...,N$ denotes the timestep index, such that $\xi_0=\tau^{-1}\int_0^\tau dt \ \xi(t)$ and $\xi_N = \tau^{-1} \int_{T-\tau}^T dt \ \xi(t)$. Using \eqref{eq:discr_path_int}, Eqs.~\eqref{eq:conv_o1} and \eqref{eq:conv_o2} reduce to
\begin{align}
    \widebarbar{\dot{x}(t)}^{(v)} P_{\dot{x}}(v) = \int d\xi_N~ P_\xi(\xi_N) \overline{w(t)}^{(v-\xi_N)} P_w(v-\xi_{N}) \label{eq:empiricalFirstMoment}\\
    \left( \widebarbar{\dot{x}^2(t)}^{(v)} - \frac{2D_x}{\tau} \right) P_{\dot{x}}(v) = \int d\xi_N~ P_\xi(\xi_N)  \overline{w^2(t)}^{(v-\xi_N)} P_w(v-\xi_N)~,\label{eq:empiricalSecondMoment}
\end{align}
which can be rearranged, via straightforward deconvolution, into expressions for the desired first- and second-order conditional moments of the hidden self-propulsion velocity $w(t)$. Note that, while the above shows such indirect measurement is possible, practical difficulties may arise, particularly when the state space of $w(t)$ is discrete. For example, deconvolution in Fourier space requires a suitable representation on the whole real line $v \in \mathbb{R}$ for the left-hand sides of Eqs.~\eqref{eq:empiricalFirstMoment} and \eqref{eq:empiricalSecondMoment}, which are obtained empirically and thus have finite support by definition. Similar problems prevent us from na{\"i}vely ``dividing through'' by the empirical probability $P_w$. The treatment of such technical difficulties constitutes an interesting problem in itself \cite{andrews2013picture}, which however lies beyond the scope of the present work.  

\subsection{Numerical verification of Eqs.~\eqref{eq:empiricalFirstMoment} and \eqref{eq:empiricalSecondMoment} for an AOU particle}
\begin{figure}
    \centering
    \includegraphics[width=0.98\textwidth, trim = 0cm 0.6cm 1.4cm 0.4cm, clip]{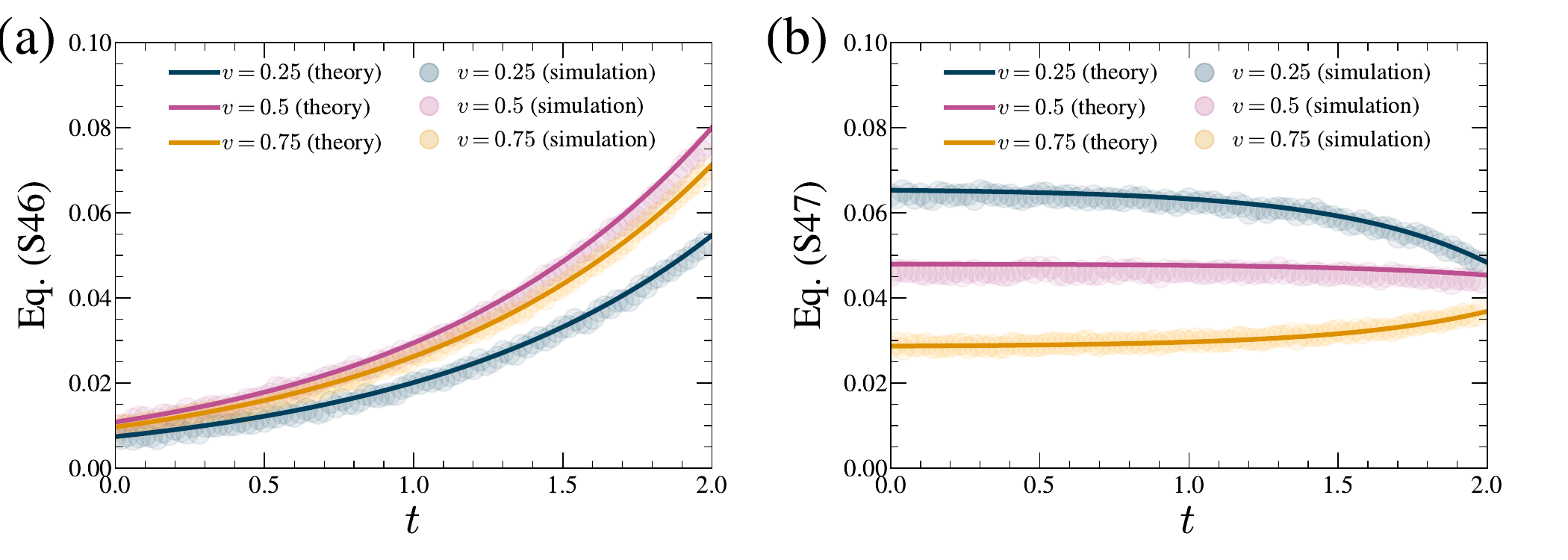}
    \caption{Numerical verification of (a) Eq.~\eqref{eq:empiricalFirstMoment} and (b) Eq.~\eqref{eq:empiricalSecondMoment} for the particular case of an AOU particle with $\mu = 1$, $D_w =0.1$, $D_x = 10^{-4}$. Solid curves indicate the analytical evaluation of the right-hand sides, as written in Eqs.~\eqref{eq:nv_46} and \eqref{eq:nv_47} respectively. Markers indicate empirical values of the left-hand sides of Eqs.~\eqref{eq:empiricalFirstMoment} and \eqref{eq:empiricalSecondMoment}, obtained from numerical simulations of AOU motion in one dimension (timestep of the numerical integration was $dt=10^{-4}$, timestep of the empirical trajectory was $\tau = 10^{-3}$ and $T=2$). }
\label{fig:numerical_test_s47-47}
\end{figure}

We verify Eqs.~(\ref{eq:empiricalFirstMoment}) and (\ref{eq:empiricalSecondMoment}) in the case of an AOU particle, as described in Sec.~\ref{sr:V}, for which the first and second conditional moments are available in closed form. The integral appearing on the right-hand side of each equation can be performed explicitly, giving 
\begin{align}
    \widebarbar{\dot{x}(t)}^{(v)} P_{\dot{x}}(v) =& \left(\frac{\mu \tau^3 D_w^2 v^2}{2 \pi (2 \mu D_x + \tau D_w)}\right)^{\frac{1}{2}}\exp\left(- \mu \left[(T-t) + \frac{v^2 \tau}{2 D_w \tau + 4 \mu D_x}\right]\right), \label{eq:nv_46}\\
    \left( \widebarbar{\dot{x}^2(t)}^{(v)} - \frac{2D_x}{\tau} \right) P_{\dot{x}}(v) =& \left(\frac{\mu \tau}{2 \pi (2 \mu D_x + \tau D_w)}\right)^{\frac{1}{2}}\Bigg[\frac{D_w}{\mu}(1 - \exp\left(- 2 \mu (T-t)\right) \exp\left(- \frac{\mu v^2 \tau}{2D_w \tau + 4 \mu D_x}\right) \nonumber \\ 
    & + \left(\left(\frac{v \tau D_w}{2\mu D_x + \tau D_w}\right)^2 + \frac{2 D_x D_w}{2\mu D_x + \tau D_w}\right) \exp\left(- \mu \left[2(T-t) + \frac{v^2 \tau}{2 D_w \tau + 4 \mu D_x}\right]\right) \Bigg].  \label{eq:nv_47}
\end{align}
The comparison of these predictions to data from numerical experiments is shown in Fig.~\ref{fig:numerical_test_s47-47}, where the right-hand sides of Eqs.~(\ref{eq:empiricalFirstMoment}) and (\ref{eq:empiricalSecondMoment}), given explicitly by Eqs.~\eqref{eq:nv_46} and \eqref{eq:nv_47}, are plotted against numerical estimates of the left-hand sides. Good agreement between the theoretical results and simulations is observed in all cases.

\section{Boundary-update protocol} \label{sr:IV}
\subsection{Derivation of conditional splitting probabilities for a run-and-tumble particle}\label{SM_sec:SplitProbDerivation}
In this supplementary section, we derive the conditional splitting probabilities that are the foundation of the numerical ``boundary-update" protocol conjectured to achieve optimal power extraction from a RnT particle for all $\rm Pe$. This generalises the known results for the (unconditional) splitting probabilities for a run-and-tumble particle \cite{malakar2018steady}.

Consider a run-and-tumble process governed by the Langevin equation $\dot{x}(t) = \nu w(t) + \sqrt{2D_x} \xi(t)$, where the normalised self-propulsion state $w \in \{-1,1\}$ follows a symmetric telegraph process with switching rate $\alpha$. We say the particle is in a right-moving (left-moving) state at time $t$ if $w(t) = +1$ ($w(t)=-1$).
Let $\Pi_{s_1 L/2}^{s_2}(x_{0}, s_3)$ with $s_{1,2,3} \in \{-,+\}$ denote the probability for a particle to exit through the boundary at $x = s_1 L/2$ in the $w = s_2$ state given it was initialised at position $x_{0} \in [-L/2,L/2]$ in the $w = s_3$ state. For example, $\Pi_{-L/2}^{-}(x_{0}, +)$ denotes the probability for a particle to exit through the left-hand boundary at $x=-L/2$ in the left-moving state given it was initialised at position $x_{0} \in [-L/2,L/2]$ in the right-moving state. Four of the eight combinations of $s_{1,2,3}$ can be readily obtained from the remaining ones by exploiting the symmetries $\Pi_{L/2}^{\pm}(x_0, +) = \Pi_{-L/2}^{\mp}(-x_0, -)$ and $\Pi_{L/2}^{\pm}(x_0, -) = \Pi_{-L/2}^{\mp}(-x_0, +)$. As such, we will calculate only the splitting probabilities to exit at the left-hand boundary, $x = -L/2$. To ease notation, we define $\pi^{\pm}(x_0, \pm) \equiv \Pi_{-L/2}^{\pm}(x_0, \pm)$.

We first derive the governing ODEs for the splitting probabilities by starting from the microscopic description on a lattice and taking the continuum limit \cite{van1992stochastic}. On a lattice of spacing $\delta$, a particle undergoes one of three processes at each time step,
\begin{enumerate}
    \item hopping to the right-adjacent site, $x + \delta$, with rate $h_{r}$,
    \item hopping to the left-adjacent site, $x - \delta$, with rate $h_{\ell}$,
    \item switching internal state with rate $\alpha$, while remaining at the current site, $x$.
\end{enumerate}
Whence, for a particle leaving through $x=-L/2$ in the $w=-1$ state, we have the identities
\begin{subequations}\label{app:eq:SplittingProbDifferenceEq}
\begin{alignat}{2}
    \pi^{-}(x,+) &= \frac{h_{r}^{+}}{h_{r}^{+} + h_{\ell}^{+} + \alpha}\pi^{-}(x+\delta,+) + \frac{h_{\ell}^{+}}{h_{r}^{+} + h_{\ell}^{+} + \alpha}\pi^{-}(x-\delta,+) + \frac{\alpha}{h_{r}^{+} + h_{\ell}^{+} + \alpha}\pi^{-}(x,-)~,\\
    \pi^{-}(x,-) &= \frac{h_{r}^{-}}{h_{r}^{-} + h_{\ell}^{-} + \alpha}\pi^{-}(x+\delta,-) + \frac{h_{\ell}^{-}}{h_{r}^{-} + h_{\ell}^{-} + \alpha}\pi^{-}(x-\delta,-) + \frac{\alpha}{h_{r}^{-} + h_{\ell}^{-} + \alpha}\pi^{-}(x,+)~,
    \end{alignat}
\end{subequations}
where, for instance, $h_{\ell}^{+}$ is the rate for a right-moving particle to hop to the left such that $h_{\ell}^{+}/(h_{r}^{+} + h_{\ell}^{+} + \alpha)$ is the transition probability for a right-moving particle to hop to the left. Expanding each term on the right-hand side of Eq.~(\ref{app:eq:SplittingProbDifferenceEq}) in small $\delta$, e.g.\
\begin{equation}\label{app:eq:SplittingProbTaylorExpansion}
    \pi^{-}(x-\delta,-) = \pi^{-}(x,-) - \delta \frac{\mathrm{d}\pi^{-}(x,-)}{\mathrm{d}x} + \frac{1}{2}\delta^{2}\frac{\mathrm{d}^{2}\pi^{-}(x,-)}{\mathrm{d}x^{2}} + \mathcal{O}(\delta^3)~,
\end{equation}
produces, after rearrangement and up to second order in $\delta$,
\begin{subequations}\label{app:eq:SplittingProbCoupledLattice}
    \begin{alignat}{2}
        0 &= \delta\left(h_{r}^{+}-h_{\ell}^{+}\right) \frac{\mathrm{d}\pi^{-}(x,+)}{\mathrm{d}x} + \frac{1}{2}\delta^{2}\left(h_{r}^{+}+h_{\ell}^{+}\right)\frac{\mathrm{d}^{2}\pi^{-}(x,+)}{\mathrm{d}x^{2}} + \alpha\left(\pi^{-}(x,-) - \pi^{-}(x,+)\right)~,\\
        0 &= \delta\left(h_{r}^{-}-h_{\ell}^{-}\right) \frac{\mathrm{d}\pi^{-}(x,-)}{\mathrm{d}x} + \frac{1}{2}\delta^{2}\left(h_{r}^{-}+h_{\ell}^{-}\right)\frac{\mathrm{d}^{2}\pi^{-}(x,-)}{\mathrm{d}x^{2}} + \alpha\left(\pi^{-}(x,+) - \pi^{-}(x,-)\right)~.
    \end{alignat}
\end{subequations}
Finally, taking $\delta \rightarrow 0$ while keeping $\nu = \delta(h_{r}^{+}-h_{\ell}^{+}) = \delta(h_{\ell}^{-}-h_{r}^{-})$ and $D_x = \delta^{2}(h_{r}^{+}+h_{\ell}^{+})/2 = \delta^{2}(h_{r}^{-}+h_{\ell}^{-})/2$ finite results in the following coupled backward equations,
\begin{subequations}\label{app:eq:SplittingProbCoupledODEs}
    \begin{alignat}{2}
        0 &= \nu \frac{\mathrm{d}\pi^{-}(x,+)}{\mathrm{d}x} + D_x \frac{\mathrm{d}^{2}\pi^{-}(x,+)}{\mathrm{d}x^{2}} + \alpha\left(\pi^{-}(x,-) - \pi^{-}(x,+)\right) \label{app:eq:SplittingProbCoupledODEsRightMover}~,\\
        0 &= -\nu\frac{\mathrm{d}\pi^{-}(x,-)}{\mathrm{d}x} + D_x \frac{\mathrm{d}^{2}\pi^{-}(x,-)}{\mathrm{d}x^{2}} + \alpha\left(\pi^{-}(x,+) - \pi^{-}(x,-)\right) \label{app:eq:SplittingProbCoupledODEsLeftMover}~.
    \end{alignat}
\end{subequations}
By a similar procedure, the backward equations for the particle exiting the interval at $x=-L/2$ as a right mover can be shown to satisfy
\begin{subequations}\label{app:eq:SplittingProbCoupledODEsExitAsRight}
    \begin{alignat}{2}
        0 &= \nu \frac{\mathrm{d}\pi^{+}(x,+)}{\mathrm{d}x} + D_x \frac{\mathrm{d}^{2}\pi^{+}(x,+)}{\mathrm{d}x^{2}} + \alpha\left(\pi^{+}(x,-) - \pi^{+}(x,+)\right) \label{app:eq:SplittingProbCoupledODEsExitAsRightRightMover}~,\\
        0 &= -\nu\frac{\mathrm{d}\pi^{+}(x,-)}{\mathrm{d}x} + D_x \frac{\mathrm{d}^{2}\pi^{+}(x,-)}{\mathrm{d}x^{2}} + \alpha\left(\pi^{+}(x,+) - \pi^{+}(x,-)\right) \label{app:eq:SplittingProbCoupledODEsExitAsRightLeftMover}~.
    \end{alignat}
\end{subequations}

Since $\pi^{-}(x,\pm)$ and $\pi^{+}(x,\pm)$ are governed by the same ODEs, we proceed to solve Eqs.~(\ref{app:eq:SplittingProbCoupledODEs}) and (\ref{app:eq:SplittingProbCoupledODEsExitAsRight}) simultaneously up till the point of applying the different boundary conditions for each case, which are
\begin{equation}\label{app:eq:SplitProbBoundaryConditions}
    \pi^{\pm}(-L/2,\pm) = 1~, \quad
    \pi^{\mp}(-L/2,\pm) = 0~, \quad
    \pi^{\pm}(L/2,\pm) = 0~, \quad
    \pi^{\mp}(L/2,\pm) = 0~, \quad
\end{equation}
i.e.\ only one boundary condition for each statistic is non-vanishing due to the added constraint on the internal state as the particle exits the interval.

We define $\rho^{\pm}(x) \equiv \pi^{\pm}(x,-) + \pi^{\pm}(x,+)$ and $\sigma^{\pm}(x) \equiv \pi^{\pm}(x,-) - \pi^{\pm}(x,+)$ such that, after adding and subtracting Eqs.~(\ref{app:eq:SplittingProbCoupledODEsRightMover}) and (\ref{app:eq:SplittingProbCoupledODEsLeftMover}) (or, equivalently, Eqs.~(\ref{app:eq:SplittingProbCoupledODEsExitAsRightRightMover}) and (\ref{app:eq:SplittingProbCoupledODEsExitAsRightLeftMover})), we obtain
\begin{subequations}\label{app:eq:SplittingProbCoupledODEsDensityPolarity}
    \begin{alignat}{2}
        0 &= \frac{\mathrm{d}^{2}\rho^{\pm}(x)}{\mathrm{d}x^{2}} - \frac{\nu}{D_x} \frac{\mathrm{d}\sigma^{\pm}(x)}{\mathrm{d}x} \label{app:eq:DensityODE}~,\\
        0 &= \frac{\mathrm{d}^{2}\sigma^{\pm}(x)}{\mathrm{d}x^{2}} - \frac{\nu}{D_x} \frac{\mathrm{d}\rho^{\pm}(x)}{\mathrm{d}x} -\frac{2\alpha}{D_x}\sigma^{\pm}(x) \label{app:eq:PolarityODE}~,
    \end{alignat}
\end{subequations}
with updated boundary conditions
\begin{equation}\label{app:eq:SplitProbDensityPolarityBoundaryConditions}
    \rho^{\pm}\left(-L/2\right) = 1~, \quad
    \sigma^{\pm}\left(-L/2\right) = \mp 1~, \quad
    \rho^{\pm}\left(L/2\right) = 0~, \quad
    \sigma^{\pm}\left(L/2\right) = 0~. \quad
\end{equation}
Integrating Eq.~(\ref{app:eq:DensityODE}),
\begin{equation}\label{app:eq:SplitProbDensityFirstOrderODE}
     \frac{\mathrm{d}\rho^{\pm}(x)}{\mathrm{d}x} = \frac{\nu}{D_x}\sigma^{\pm}(x) - \frac{D_x}{\nu}k^{2}c_{1}^{\pm}~,
\end{equation}
and substituting the right-hand side of Eq.~(\ref{app:eq:SplitProbDensityFirstOrderODE}) in place of $\rho'(x)$ in Eq.~(\ref{app:eq:PolarityODE}), yields
\begin{equation}\label{app:eq:SplitProbPolaritySecondOrderODE}
     \frac{\mathrm{d}^{2}\sigma^{\pm}(x)}{\mathrm{d}x^{2}} = k^{2} ( \sigma^{\pm}(x) - c_{1}^{\pm})~,
\end{equation}
which has the general solution
\begin{equation}\label{app:eq:SigmaSolution}
     \sigma^{\pm}(x) = c_{3}^{\pm}e^{kx} + c_{2}^{\pm}e^{-kx} + c_{1}^{\pm}~,
\end{equation}
where $k = \sqrt{\nu^{2}/D_x^{2} + 2\alpha/D_x} > 0$ and the $c_{i}^{\pm}$ are constants of integration. Substituting the solution for $\sigma^{\pm}(x)$, Eq.~(\ref{app:eq:SigmaSolution}), into Eq.~(\ref{app:eq:SplitProbDensityFirstOrderODE}) and integrating leads to
\begin{equation}\label{app:eq:RhoSolution}
     \rho^{\pm}(x) = \frac{\nu}{D_x k} \left( c_{4}^{\pm} + c_{3}^{\pm}e^{kx} - c_{2}^{\pm}e^{-kx}\right) - \frac{2\alpha}{\nu} c_{1}^{\pm} x ~.
\end{equation}
Now, applying the boundary conditions for the separate $w = \pm 1$ cases, Eq.~(\ref{app:eq:SplitProbDensityPolarityBoundaryConditions}), to Eqs.~(\ref{app:eq:SigmaSolution}) and (\ref{app:eq:RhoSolution}) fixes the constants of integration as
\begin{subequations}\label{app:eq:SplittingProbConstants}
    \begin{alignat}{4}
        c_{1}^{\pm} &= \frac{1}{2}\frac{\cosh(\frac{kL}{2}) \mp \frac{\nu}{D_x k}\sinh(\frac{kL}{2})}{\frac{\alpha L}{\nu}\cosh(\frac{kL}{2}) + \frac{\nu}{D_x k}\sinh(\frac{kL}{2})}~,\\
        c_{2}^{\pm} &= \mp \frac{1}{4}\left( \frac{\frac{\alpha L}{\nu} \pm 1}{\frac{\alpha L}{\nu} \cosh(\frac{kL}{2}) + \frac{\nu}{D_x k}\sinh(\frac{kL}{2})} + \mathrm{csch}\left(\frac{kL}{2}\right) \right)~,\\
        c_{3}^{\pm} &= \mp \frac{1}{4}\left(\frac{\frac{\alpha L}{\nu} \pm 1}{\frac{\alpha L}{\nu} \cosh(\frac{kL}{2}) + \frac{\nu}{D_x k}\sinh(\frac{kL}{2})} - \mathrm{csch}\left(\frac{kL}{2}\right) \right)~,\\
        c_{4}^{\pm} &= \frac{1}{2} \left( \frac{D_x k}{\nu} \mp \coth\left(\frac{kL}{2}\right) \right)~,
    \end{alignat}
\end{subequations}
such that $\pi^{\pm}(x_0,-) = (\rho^{\pm}(x_0) + \sigma^{\pm}(x_0))/2$ and $\pi^{\pm}(x_0,+) = (\rho^{\pm}(x_0) - \sigma^{\pm}(x_0))/2$ are now fully determined by Eqs.~(\ref{app:eq:SigmaSolution})-(\ref{app:eq:SplittingProbConstants}).

\begin{figure}
\subfloat[Probabilities to observe a left mover, $w=-1$, at the left-hand boundary, $x=-L/2$. \label{subfig:LeftMoverAtLef}]{%
  \includegraphics[width=0.327\textwidth, trim = 13.9cm 0.3cm 15.5cm 0.2cm, clip]{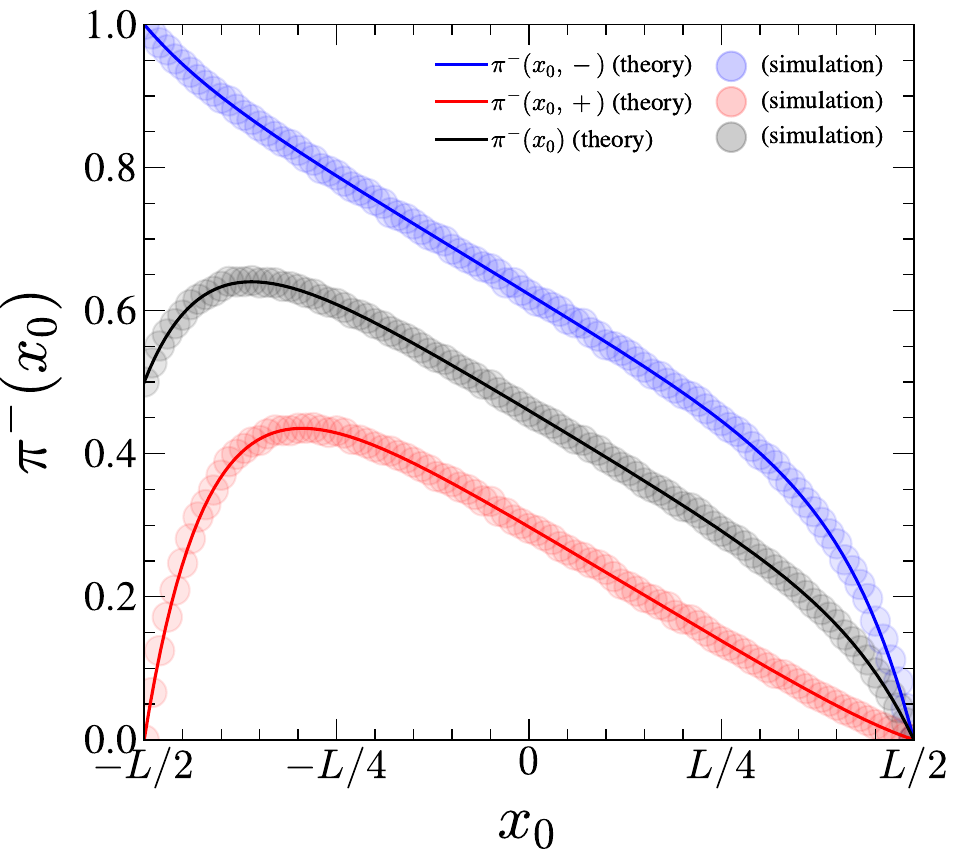}%
}\hfill
\subfloat[Probabilities to observe a right mover, $w=+1$, at the left-hand boundary, $x=-L/2$. \label{subfig:RightMoverAtLeft}]{%
 \includegraphics[width=0.33\textwidth, trim = 13.9cm 0.6cm 15.6cm 0.3cm, clip]{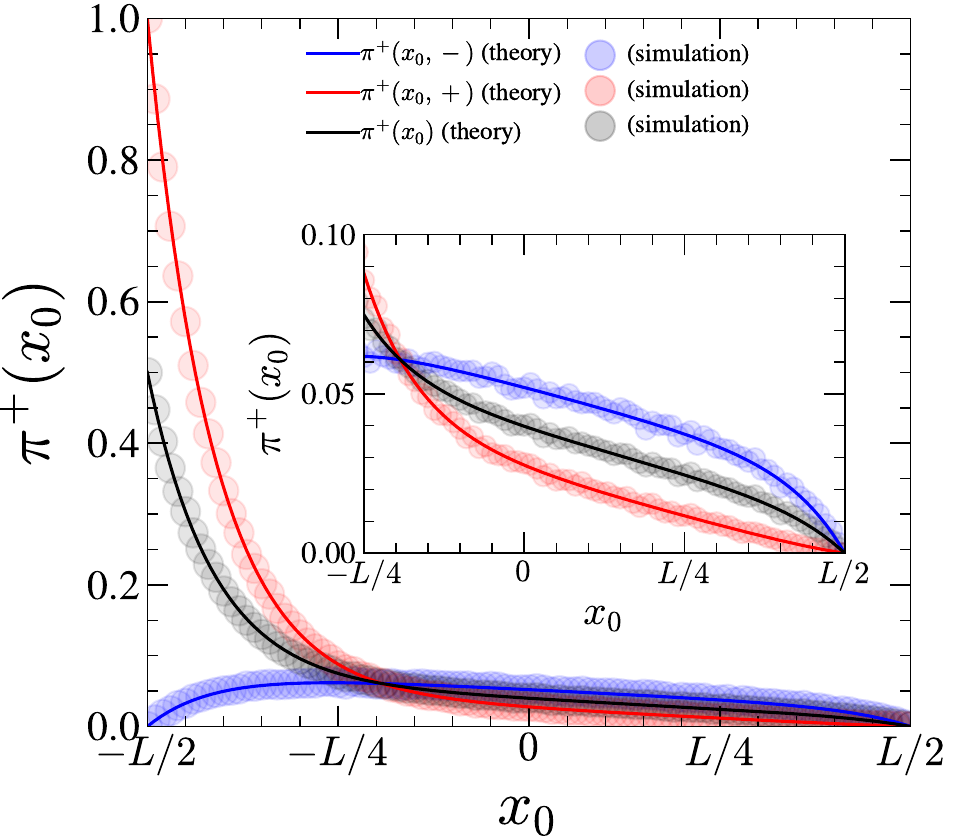}%
}\hfill
\subfloat[Probabilities to observe a particle at the left-hand boundary, $x=-L/2$. \label{subfig:AnyParticleAtLeft}]{%
  \includegraphics[width=0.33\textwidth, trim = 13.9cm 0.6cm 15.6cm 0.3cm, clip]{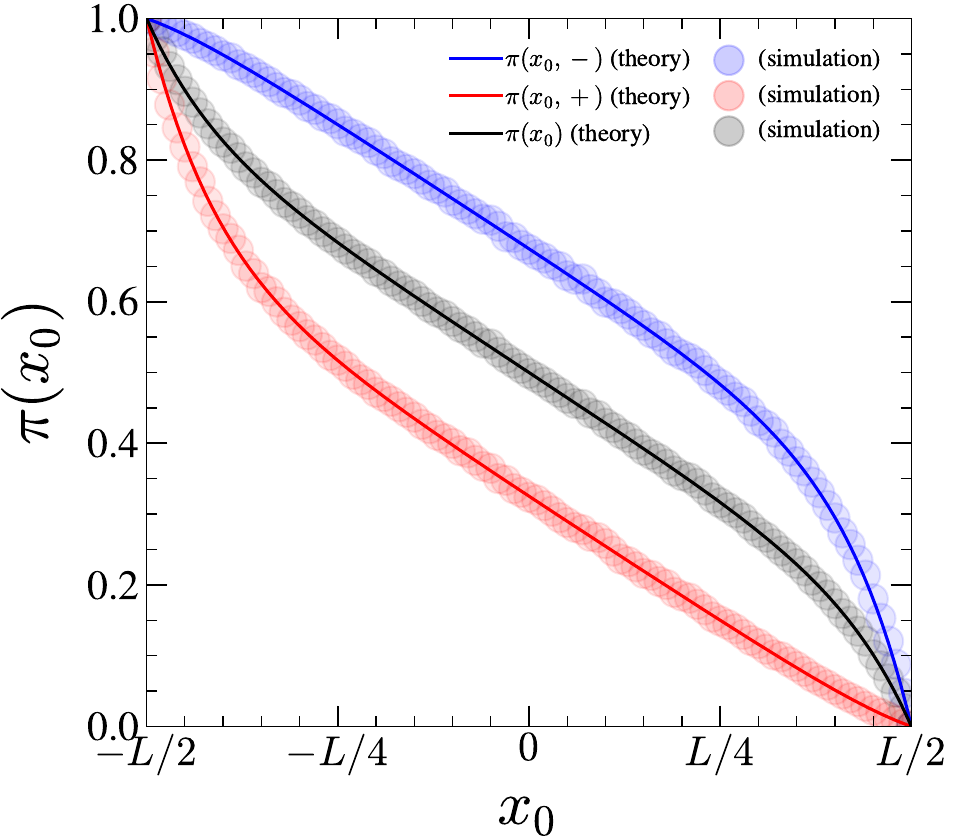}%
}
\caption{Splitting probabilities, as a function of initialisation position $x_{0}$, for a RnT particle with $\nu=1$, $D_x = 0.1$ and $\alpha = 2$, to cross the left-hand boundary at $x=-L/2$, before crossing the right-hand boundary at $x=L/2$, if it exits as (a) a left mover, (b) a right mover, or (c) either a left mover or a right mover. In each subfigure, red (respectively, blue/black) lines indicate the initialisation of a right mover (respectively, left mover/equal superposition of a left mover and a right mover). Subfigure (c) corresponds to the (unconditional) splitting probabilities derived in Ref.~\cite{malakar2018steady}. Monte-Carlo simulations (markers) were performed by numerically integrating the Langevin equation, $\dot{x} = \nu w(t) + \sqrt{2D_{x}}\xi(t)$, in timesteps of $\Delta t = 10^{-5}$ to determine the proportion of times the particle exits through the left-hand boundary for $10^{5}$ realisations at each $x_{0}/L = 0, 0.01, \dots, 1$. The theoretical results are in good agreement with simulations.}\label{fig:SplittingProbabilities}
\end{figure}

The conditional splitting probabilities, which we verify with Monte-Carlo simulations, are plotted in Fig.~\ref{fig:SplittingProbabilities}. Interestingly, we observe that there can be a higher likelihood
to observe a right mover at the left-hand boundary if it is first initialised as a left mover (Fig.~\ref{subfig:RightMoverAtLeft} in the range $x_0 \gtrapprox -L/4$). Such a phenomenon can be studied only through the conditional splitting probabilities derived here. By summing suitable combinations of the conditional splitting probabilities, i.e.\ $\pi(x_0,\pm) = \pi^{-}(x_0,\pm) + \pi^{+}(x_0,\pm)$, where $\pi(x_0,\pm)$ is the probability to observe \emph{any} particle at the left-hand boundary given initialisation at $x_{0}$ in the $w = \pm 1$ state, we can verify our results against the unconditional splitting probabilities derived in Ref.~\cite{malakar2018steady}, see Fig.~\ref{subfig:AnyParticleAtLeft}.

Finally, to establish the foundation of the boundary-update protocol mentioned in the main text, one can use Bayes' theorem in combination with the splitting probabilities derived above to determine the posterior likelihood of a particle's internal state, $w(T) = \pm 1$, \emph{given} it has crossed a particular boundary at time $t = T$. For instance, the probability of a particle being a left mover given it was initialised at $x_0=0$ and has crossed the left-hand boundary at $x=-L/2$ is
\begin{subequations}\label{app:eq:SplittingProb_baesrem}
\begin{equation}\label{app:eq:BayesianSplitUpdate}
    \begin{split}
        \mathbb{P}[w(T) = -1 | \text{left exit}, x_0 = 0 ]
        &= \frac{\mathbb{P}[w(T) = -1 \cap \text{left exit} ]}{ \mathbb{P}[\text{left exit} ] }\\
        &= \frac{ \pi^{-}(0,-)\mathbb{P}[w(0) = -1] + \pi^{-}(0,+)\mathbb{P}[w(0) = +1] }{ \pi(0,-) \mathbb{P}[w(0) = -1] + \pi(0,+) \mathbb{P}[w(0) = +1] }~,
     \end{split}
\end{equation}
where $\mathbb{P}[w(0) = \pm1]$ denotes the prior probability of the self-propulsion mode at initialisation. Similarly,
    \begin{alignat}{3}
    \mathbb{P}[w(T) = -1 | \text{right exit}, x_0 = 0 ]
        &= \frac{ \Pi^{-}_{L/2}(0,-)\mathbb{P}[w(0) = -1] + \Pi^{-}_{L/2}(0,+)\mathbb{P}[w(0) = +1] }{ (1-\pi(0,-)) \mathbb{P}[w(0) = -1] + (1-\pi(0,+)) \mathbb{P}[w(0) = +1] }~, \\
    \mathbb{P}[w(T) = +1 | \text{left exit}, x_0 = 0 ]
        &= \frac{ \pi^{+}(0,-)\mathbb{P}[w(0) = -1] + \pi^{+}(0,+)\mathbb{P}[w(0) = +1] }{ \pi(0,-) \mathbb{P}[w(0) = -1] + \pi(0,+) \mathbb{P}[w(0) = +1] }~, \label{eq: rightmoverleftexit}\\
    \mathbb{P}[w(T) = +1 | \text{right exit}, x_0 = 0 ]
        &= \frac{ \Pi^{+}_{L/2}(0,-)\mathbb{P}[w(0) = -1] + \Pi^{+}_{L/2}(0,+)\mathbb{P}[w(0) = +1] }{ (1-\pi(0,-)) \mathbb{P}[w(0) = -1] + (1-\pi(0,+)) \mathbb{P}[w(0) = +1] } ~. \label{eq: rightmoverrightexit}
    \end{alignat}
\end{subequations}

\subsection{Numerical implementation of boundary-update protocol}
In this supplementary section we describe 
the numerical implementation of the boundary-update protocol for the RnT dynamics discussed in the main text. The boundary-update protocol draws on Eq.~(\ref{eq:opt_force_general}) in the main text, which states that the optimal protocol $F_{\rm ext}^*(t)$ is proportional to the instantaneous posterior expectation of the hidden self-propulsion velocity at time $t$ given the entire observed trajectory of the velocity, $\dot{x}(t')$ for $t' \in [0,t]$. 
In this scheme, the posterior expectation is  updated on the fly upon the RnT particle exiting an interval of interest using the conditional splitting probabilities in Eq.~\eqref{app:eq:SplittingProb_baesrem}. 

In the following, $p_{\rm prior}(t)$ and $p_{\rm post}(t)$ denote the Bayesian prior and posterior probability respectively (with respect to the observation that the particle has exited the interval) that the particle is currently a right-mover. Whilst the particle is in the bulk of the interval, no additional information is collected to improve the Bayesian inference of the hidden self-propulsion velocity and the posterior probability evolves in accordance to the Markov master equation 
\begin{align}
    \frac{d \mathbb{P}_w[w(t)=+1]}{dt} &= \alpha (\mathbb{P}_w[w(t)=-1] - \mathbb{P}_w[w(t)=+1]) \nonumber \\
    &= \alpha (1 - 2 \mathbb{P}_w[w(t)=+1]) ~,
\end{align}
which describes self-propulsion velocity reversal with Poisson rate $\alpha$. Accordingly,
$\dot{p}_{\rm post}(t) = \alpha (1 - 2 p_{\rm post}(t))$ which has $p_{\rm post} = 1/2$ as an attractive fixed point, reflecting the gradual loss of knowledge of the particle's internal state as time progresses. The following pseudocode summarises the numerical implementation of the protocol:

\begin{figure}{%
  \includegraphics[width=0.6\textwidth, trim = 0.1cm 0.2cm 0.1cm 0.1cm, clip]{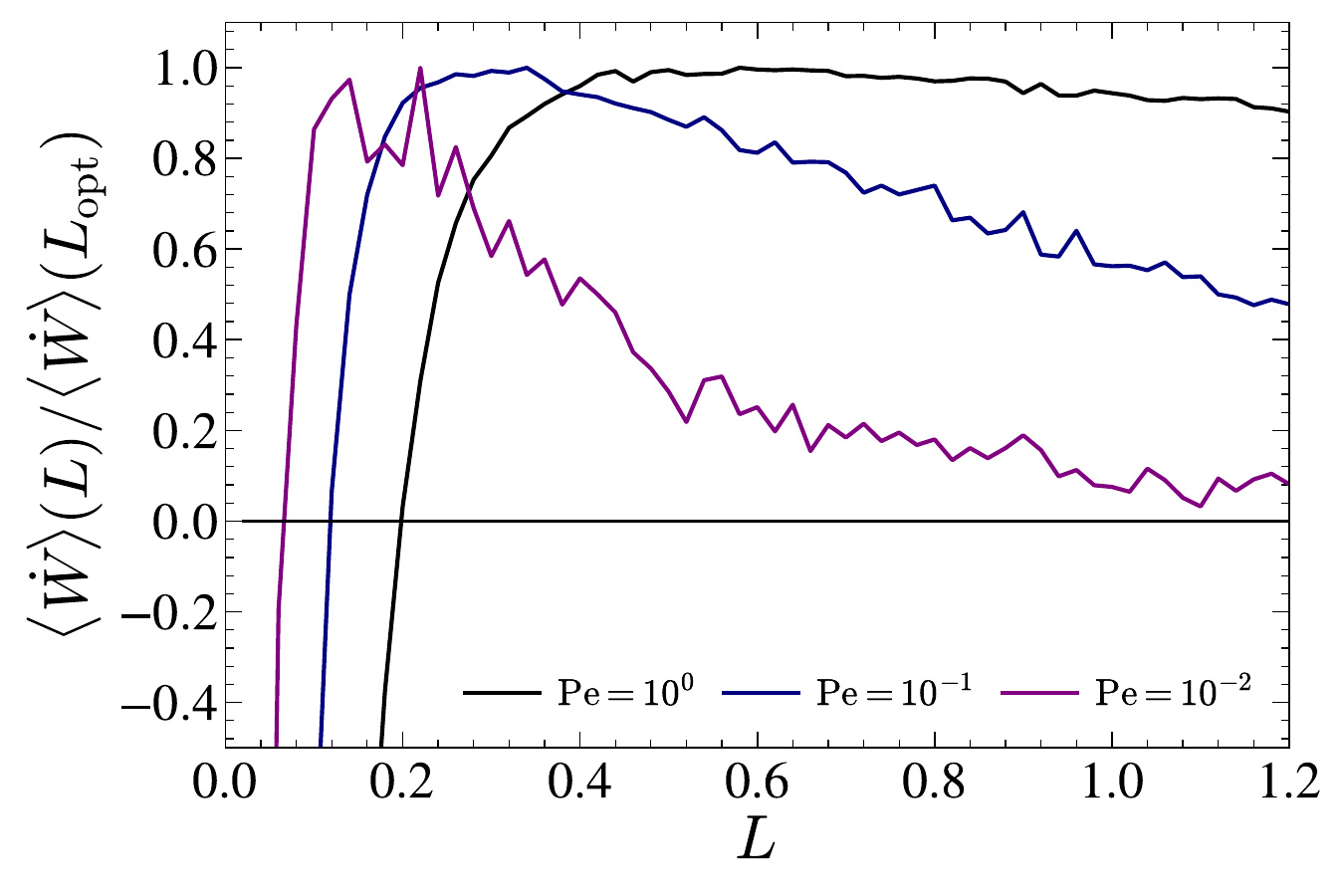}%
}
\caption{Average extracted power $\langle \dot{W} \rangle(L)$ as a function of interval length $L$ for three values of $\rm Pe$. The notation $\langle \bullet \rangle$ denotes a time-average over an entire simulated trajectory of duration $T = 3\cdot 10^5$. The extracted power is normalised by that attained using the optimal interval length $L_{\rm opt}$ at each $\rm Pe$ value. Since maximum power extraction is attained at different interval lengths $L_{\rm opt}$ for each value of $\rm Pe$, no single choice of $L$ can be used for the entire range of values of $\rm Pe$.}\label{fig:WorkvsL}
\end{figure}

\SetKwInput{Kw}{Initialise}
\begin{algorithm}[H]
 \Kw{ \\
 \quad  $ \bullet \ $ Set $t=0$, $x(0)=0$, $F_{\rm ext}(0)=0$; \\ 
 \quad  $ \bullet \ $ Set $p_{\rm prior}(0) = p_{\rm post}(0) = \mathbb{P}_w[w(0) = +1] =  \frac{1}{2}$ and $[x_-, x_+] = [-L/2, L/2]$;}

 \While{$t<T$}{
  $\bullet \ $Increment particle position according to the forward Euler scheme, i.e.\
  $x(t + \Delta t) = x(t) + \Delta t (\nu w(t) + F_{\mathrm{ext}}(t)) + \sqrt{2 D_x \Delta t}\Delta \xi$, where $\Delta \xi$ represents a unit variance Gaussian random variable drawn at each time step\;
  $\bullet \ $ Increment the extracted work by $W(t + \Delta t) = W(t) - F_{\mathrm{ext}}(t+ \Delta t) \cdot (x(t+\Delta t) - x(t))$\;
  $\bullet \ $ Increment the posterior probability according to $p_{\rm post}(t + \Delta t) = p_{\rm post}(t) +\Delta t \cdot \alpha ( 1 - 2 p_{\rm post}(t))$\;
  $\bullet \ $ Update $F_{\rm ext}(t+\Delta t) = - \nu p_{\rm post}(t+ \Delta t)/2$ in accordance with Eq.~(\ref{eq:opt_force_general})\;

  \If{$x(t+\Delta t)>x_+$ }{
   $\bullet \ $ Set posterior probability
   $p_{\rm post}(t+\Delta t) = \mathbb{P}_w[ w(t+\Delta t) = +1 | \text{right exit}]$ from  Eq.~\eqref{eq: rightmoverrightexit}, with $\mathbb{P}[w(0)=+1] = p_{\rm prior}$ and correspondingly $\mathbb{P}[w(0)=-1] = 1-p_{\rm prior}$ \; 
   $\bullet \ $ Update $F_{\rm ext}(t+\Delta t)$ according to Eq.~(\ref{eq:opt_force_general})\;
   $\bullet \ $ Update the prior probability $p_{\rm prior}(t + \Delta t) = p_{\rm post}(t + \Delta t)$\;
   $\bullet \ $ Reset boundaries to $[x_-, x_+] = [x(t+ \Delta t) - L/2, x(t+ \Delta t) + L/2]$;
   }

   \ElseIf{$x(t+ \Delta t)<x_-$}{
   $\bullet \ $ Set posterior probability
   $p_{\rm post}(t+\Delta t) = \mathbb{P}_w[ w(t+\Delta t) = +1 | \text{left exit}]$ from  Eq.~\eqref{eq: rightmoverleftexit}, with $\mathbb{P}[w(0)=+1] = p_{\rm prior}$ and correspondingly $\mathbb{P}[w(0)=-1] = 1-p_{\rm prior}$ \; 
   $\bullet \ $ Update $F_{\rm ext}(t+\Delta t)$ according to Eq.~(\ref{eq:opt_force_general})\;
   $\bullet \ $ Update the prior probability $p_{\rm prior}(t + \Delta t) = p_{\rm post}(t + \Delta t)$\;
   $\bullet \ $ Reset boundaries to $[x_-, x_+] = [x(t+ \Delta t) - L/2, x(t+ \Delta t) + L/2]$;}
   $\bullet \ $ Increment the simulation time, $t = t + \Delta t $
}
\end{algorithm}

We simulated this protocol for $M=400$ values of ${\rm Pe} = \nu^2 / (\alpha D_x)$, evenly distributed in logarithmic space in the range $10^{-3}<\rm Pe < 10^4$, and for a run duration of $T=3\cdot 10^5$ with $\Delta t = 10^{-4}$. Without loss of generality, we set $\nu =1, D_{x} = 1$, such that ${\rm Pe} = \alpha^{-1}$, and varied $\alpha$ in the range $10^{3}>\alpha > 10^{-4}$. 
Although we conjecture that optimal work extraction is achieved only in the limit $L \to 0$, i.e.\ that $\lim_{L \to 0} p_{\rm post}(t) = \mathbb{P}[w(t)|\{x\}_0^t]$, in practice the restriction of a finite simulation timestep $\Delta t$ prevents us from being able to access this limit. Instead, we found numerically that, for a given $\Delta t$, the power output is maximal at a small but finite interval size $L_{\mathrm{opt}}(\Delta t)$, that depended on $\rm Pe$. To account for this effect, the boundary-update protocol was executed for 60 evenly spaced values of $L$ in the range $L \in [0.02,1.2]$ to identify $L_{\rm opt}$. Figure~\ref{fig:WorkvsL} illustrates the variation in average extracted power $\langle \dot{W} \rangle$ with $L$ for different values of $\rm Pe$ and demonstrates that optimal work extraction cannot be attained from a single choice of interval length $L$ for all values of $\rm Pe$.

The numerical results for this protocol, which are shown in Fig.~2 of the main text, are in excellent agreement with the theoretical asymptotes derived in Eqs.~(\ref{eq:opt_w_colorful}) and (\ref{eq:RnT_lowPe_asym}). Extending this approach to continuous-state self-propulsion dynamics, e.g.\ the AOU process, remains an open challenge.
\end{document}
%